\documentclass[english]{article}
\usepackage[T1]{fontenc}
\usepackage[latin9]{inputenc}
\usepackage{booktabs}
\usepackage{amsmath}
\usepackage{amsthm}
\usepackage{amssymb}
\usepackage{graphicx}

\makeatletter

\providecommand{\tabularnewline}{\\}

\numberwithin{equation}{section}
\numberwithin{figure}{section}

\usepackage{babel}
\usepackage{fullpage}
\usepackage{authblk}
\usepackage{hyperref}
\usepackage{placeins}
\usepackage{hyphenat}
\usepackage{fancyvrb}
\usepackage[hypcap]{caption}
\usepackage{bbm} 
\usepackage{relsize}
\usepackage{amsbsy}
\usepackage{textcomp}

\newcommand{\ind}{ \mathlarger{\mathlarger{\mathbbm{1}}}}

\title{Enhancing Valuation of Variable Annuities  in Lévy Models with Stochastic Interest Rate}
\author{ \textsc{Ludovic Goudenège}\thanks{F\'ederation de Math\'ematiques de CentraleSupélec - CNRS FR3487, France -\texttt{ ludovic.goudenege@math.cnrs.fr}} 
\and \textsc{Andrea Molent}\thanks{Dipartimento di Scienze Economiche e Statistiche, Universit\`a degli Studi di Udine, Italy - \texttt{andrea.molent@uniud.it}} 
\and \textsc{Xiao Wei}\thanks{China Institute for Actuarial Science \& School of Insurance, Central University of Finance and Economics, China - \texttt{weixiao@cufe.edu.cn}} 
\and \textsc{Antonino Zanette}\thanks{Dipartimento di Scienze Economiche e Statistiche, Universit\`a degli Studi di Udine, Italy - \texttt{antonino.zanette@uniud.it}}}
\date{}

\@ifundefined{showcaptionsetup}{}{%
 \PassOptionsToPackage{caption=false}{subfig}}
\usepackage{subfig}
\makeatother

\usepackage{babel}
\begin{document}
\maketitle

\begin{flushleft}
\rule{1\columnwidth}{1pt}
\par\end{flushleft}

\begin{flushleft}
\textbf{\large{}Abstract}{\large\par}
\par\end{flushleft}

This paper extends the valuation and optimal surrender framework for
variable annuities with guaranteed minimum benefits in a Lévy equity
market environment by incorporating a stochastic interest rate described
by the Hull-White model. This approach frames a more dynamic and realistic
financial setting compared to previous literature. We exploit a robust
valuation mechanism employing a hybrid numerical method that merges
tree methods for interest rate modeling with finite difference techniques
for the underlying asset price. This method is particularly effective
for addressing the complexities of variable annuities, where periodic
fees and mortality risks are significant factors. Our findings reveal
the influence of stochastic interest rates on the strategic decision-making
process concerning the surrender of these financial instruments. Through
comprehensive numerical experiments, and by comparing our results
with those obtained through the Longstaff-Schwartz Monte Carlo method,
we illustrate how our refined model can guide insurers in designing
contracts that equitably balance the interests of both parties. This
is particularly relevant in discouraging premature surrenders while
adapting to the realistic fluctuations of financial markets. Lastly,
a comparative statics analysis with varying interest rate parameters
underscores the impact of interest rates on the cost of the optimal
surrender strategy, emphasizing the importance of accurately modeling
stochastic interest rates.

\vspace{2mm}

\noindent \emph{\large{}Keywords}: Variable Annuity; Guaranteed Minimum
Benefit; stochastic interest rate; Lévy process; tree methods; finite
difference;

\noindent\rule{1\columnwidth}{1pt}

\newpage

\section{Introduction}

In the evolving field of retirement planning and investment strategies,
variable annuities, also known as equity-linked or equity-indexed
annuities, have emerged as pivotal instruments. These products uniquely
blend the growth potential of equity investments with the security
of insurance guarantees, notably through features such as the Guaranteed
Minimum Accumulation Benefit (GMAB) and the Guaranteed Minimum Death
Benefit (GMDB). Specifically, the GMAB ensures that policyholders
are guaranteed a minimum return on their investments at the contract's
maturity, while the GMDB provides a safety net in the event of the
policyholder's untimely demise, ensuring that a predefined benefit
is payable to their beneficiaries.

However, accurately valuing these complex financial instruments requires
sophisticated mathematical models that can capture the inherent risks
and uncertainties of the equity markets. Traditional models often
fall short by assuming smooth, continuous market dynamics, thereby
overlooking the abrupt, significant shifts that characterize real-world
markets. Herein lies the importance of employing jump models, which
are more effective at simulating the market's discrete, often volatile
nature. Such models are crucial for understanding the pricing and
risk management of variable annuities with GMAB and GMDB riders, where
market volatility directly influences the contract's value and the
insurer's liability. In recent years, the valuation of variable annuities
within the framework of Lévy processes has garnered substantial attention,
highlighting the complexity and the necessity for advanced modeling
techniques in financial mathematics. To this aim, one of the first
contributions is due to Jaimungal and Young \cite{jaimungal2005pricing},
who pioneere the integration of optimal investment strategies in equity-linked
pure endowments contracts under a finite variation Lévy process by
investigating stochastic control methods. Ballotta \cite{ballotta2010efficient}
contributes by examining the pricing of ratchet equity-indexed annuities
by exploiting Fourier-transform type techniques within Lévy models.
Gerber et al. \cite{gerber2013valuing} introduce closed-form solutions
for valuing equity-linked death benefits in variable annuities under
jump diffusion models, significantly advancing actuaries' capabilities
in accurately pricing these complex insurance products. Zhou and Wu
\cite{zhou2015valuing} explore the valuation of equity-linked investment
products with a threshold expense when the underlying fund evolves
as a jump diffusion process. Hieber \cite{hieber2017cliquet} advances
the computational aspects of pricing of equity-linked life insurance
contracts that offer an annually guaranteed minimum return, by employing
efficient numerical methods based on the Fourier transform. Cui et
al. \cite{cui2017equity} delved into the design of equity-indexed
annuities and their guarantees under various market conditions and
develop a novel and efficient transform-based method to price equity-linked
annuities. Zhang et al. \cite{zhang2017levy} focus on equity-linked
death benefits using a projection method and Fast Fourier Transform
within general exponential Lévy models, offering a novel approach
for accurate valuations. Dong et al. \cite{dong2019willow} introduce
willow tree algorithms for pricing Guaranteed Minimum Withdrawal Benefits
under jump-diffusion and CEV models, significantly reducing computational
time while maintaining accuracy, and include optimal dynamic withdrawal
strategies. Zhang et al. \cite{zhang2020valuing} present an innovative
valuation method for equity-linked death benefits within general exponential
Lévy models by employing a projection method alongside Fast Fourier
Transform. 

Recently, Kirkby and Aguilar \cite{kirkby2023valuation} develop a
comprehensive framework for the valuation and optimal surrender of
equity-linked variable annuities, considering both guaranteed minimum
accumulation and death benefits within a Lévy-driven market model.
They introduce discrete-time treatment with periodic premiums and
fees, a novel approach that aligns with practical market operations,
significantly advancing the understanding and management of surrender
behaviors in variable annuity contracts.

Nevertheless, these studies collectively underscore the critical role
of incorporating jumps and stochastic components in accurately valuing
variable annuities, providing a richer understanding of the underlying
risks and market behaviors. Nonetheless, despite these significant
advancements in understanding and valuing equity-linked variable annuities,
none of the mentioned authors considers a stochastic interest rate
model. This omission represents a critical gap, given the long maturity
of the products in question. The evaluation, hedging, and lapse strategy
description fundamentally rely on an accurate interest rate model
to reflect the long-term financial landscape, making the incorporation
of stochastic interest rates an essential direction for future research
in this area. Indeed, Kirkby and Aguilar \cite{kirkby2023valuation}
highlight in their work the need for future development to incorporate
stochastic interest rates into their model, acknowledging this as
a critical aspect for further enhancing the accuracy and relevance
of their valuation framework for equity-linked variable annuities,
especially given the significant impact of interest rate movements
over time. This consideration for future research underscores the
importance of developing a comprehensive model that can capture more
precisely the complexities and dynamics of financial markets.

Building on the foundational work of Kirkby and Aguilar \cite{kirkby2023valuation},
our study introduces a critical enhancement: the incorporation of
a stochastic interest rate model, as described by the Hull-White model.
Considering a stochastic interest rate model is crucial for evaluating
variable annuity products with long maturities, such as the 25-year
products discussed in \cite{kirkby2023valuation}. Over such an extended
period, interest rates are likely to experience significant fluctuations,
which can substantially affect the product's value, its associated
guarantees, and the policyholder's behavior. A stochastic interest
rate model captures the uncertainty and dynamics of interest rates
over time, allowing for a more accurate assessment of the product's
risks and rewards. It reflects the real financial environment's complexity,
providing a comprehensive framework for evaluating the long-term financial
obligations and potential returns of variable annuities. This approach
is essential for ensuring that pricing, hedging, and risk management
strategies are robust and aligned with the evolving economic landscape.

Recently, some authors have begun to consider the problem of valuing
derivative instruments in Lévy stochastic models that also include
the stochastic interest rate. Boyarchenko et al. \cite{boyarchenko2017efficient}
delve into option pricing under Lévy processes without specifying
the interest rate model, thus offering a broad framework applicable
across different stochastic interest rate environments. Bao and Zhao
\cite{bao2019option} consider European option pricing in Markov-modulated
exponential Lévy models, where stochastic interest rates are modeled
by a Markovian regime-switching Hull-White process, showcasing a novel
integration of regime switching into the stochastic interest rate
context. Tan et al. \cite{tan2020pricing} extend traditional jump-diffusion
models to include general Lévy processes with stochastic interest
rates, specifically employing the Girsanov theorem and Itô formula
for pricing European-style options, without detailing the exact interest
rate model used. As far as variable annuities are considered, regime-switching
models are investigated by Costabile \cite{costabile2017lattice}
who propose a lattice-based model to evaluate contract than embody
guaranteed minimum withdrawal benefits.

Our approach acknowledges the complex interplay between equity market
jumps and interest rate variability, necessitating a robust numerical
method capable of handling this multidimensional challenge. To this
end, we draw inspiration from the hybrid numerical method developed
by Briani et al. \cite{briani2019numerical}, originally designed
for options pricing under stochastic volatility and stochastic interest
rate. Such a method integrates a hybrid approximation approach for
option pricing in the Bates model with stochastic interest rate, using
a blend of tree methods for volatility and interest rate dimensions,
and finite difference techniques for the underlying asset price. This
framework is notably flexible and efficient for handling the complexities
introduced by stochastic interest rates, as it allows for the accurate
modeling of the interest rate's impact on option pricing through the
application of a finite difference scheme to a partial integro-differential
equation (PIDE). To address this PIDE, multiple strategies are available,
among which is the Implicit-Explicit (IMEX) approach developed by
Cont and Voltchkova \cite{cont2005finite} or the Wiener-Hopf approximated
factorization as investigated by Kudryavtsev \cite{Kudryavtsev2011}. 

This paper aims to bridge the gap between theoretical financial models
and the practical exigencies of the variable annuities market. By
integrating a stochastic interest rate into the valuation model introduced
by Kirkby and Aguilar \cite{kirkby2023valuation} and employing an
advanced numerical solution, we provide insights that are not only
theoretically rigorous but also immensely valuable for practitioners
dealing with the intricacies of GMAB and GMDB contracts. Our work
underscores the indispensable role of sophisticated modeling and numerical
techniques in the actuarial science and financial engineering domains,
especially as they pertain to products that play a crucial role in
individuals' financial security and retirement planning. Finally,
we stress out that the hybrid nature of the method introduced by Briani
et al. \cite{briani2019numerical} facilitates the handling of the
multi-faceted dynamics present in variable annuities with GMAB and
GMDB features, providing a robust tool for their valuation in a market
environment where both equity jumps and interest rate fluctuations
are critical factors. Moreover, this method is also particularly interesting
in that it is very versatile and can be used for various Lévy processes.

In our study, we compare the proposed Hybrid method with a  Monte
Carlo method based on the Longstaff-Schwarz algorithm, which has been
adapted to handle the challenging task of learning the continuation
value in scenarios characterized by abrupt value shifts and significant
convexity changes. Our analysis spans four Lévy models, namely the
Normal Inverse Gaussian, the Variance Gamma, the CGMY, and the Merton
Jump Diffusion, revealing that both numerical approaches yield consistent
valuations, with the hybrid method demonstrating rapid convergence.
Focused tests on calculating the surrender premium, which is critical
for understanding lapse dynamics, are supplemented by a qualitative
analysis. Through comparative statics analysis, we explore how the
surrender premium varies within this stochastic rate model and examine
the impact of rate parameters on the surrender premium, offering deeper
insights into its behavior and the factors influencing it.

The remainder of the paper is organized as follows. In Section 2 we
frame the stochastic model. In Section 3 we describe the insurance
contract and the principles useful in its evaluation. In Section 4
we present the pricing algorithms. In Section 5 we discuss numerical
results. Finally, in Section 6, we conclude.

\section{The Lévy models with stochastic interest rate}

Lévy processes represent a fundamental class of stochastic processes
with wide-ranging applications in finance, particularly in the modeling
of asset returns and the pricing of derivatives. These processes are
characterized by their capability to incorporate jumps, thereby capturing
the discontinuities and heavy tails often observed in financial data. 

A Lévy process is a type of stochastic process characterized by having
stationary and independent increments (for general definitions, see,
\emph{e.g.}, Sato \cite{sato1999levy}). Notably, a Lévy process may
encompass a Gaussian component, a pure jump component, or both. The
latter is characterised by the density of jumps, which is termed the
Lévy density. Now, let us consider a filtered risk-neutral probability
space $(\Omega,\mathbb{Q},\{\mathcal{F}_{t}\}_{t\geq0})$ under which
valuations are made, and in particular $\mathbb{E}\left[\cdot\right]=\mathbb{E}^{\mathbb{Q}}\left[\cdot\right]$.
Then, a Lévy process $X$ can be completely specified by its characteristic
exponent, $\psi$, definable from the equality $\mathbb{E}\left[e^{i\xi X(t)}\right]=e^{-t\psi(\xi)}$.
The characteristic exponent $\psi$ is described by the Lévy-Khintchine
formula, which reads
\begin{equation}
\psi(\xi)=\frac{\sigma^{2}}{2}\xi^{2}-i\mu\xi+\int_{\mathbb{R}\setminus\left\{ 0\right\} }\left(1-e^{i\xi y}+i\xi y\mathbf{1}_{|y|\leq1}\right)\nu(dy),\label{psi1}
\end{equation}
where $\sigma^{2}\ge0$ is the variance of the Gaussian component
and the Lévy measure $\nu(dy)$ satisfies the following relation:
\begin{equation}
\int_{\mathbb{R}\setminus\left\{ 0\right\} }\min\{1,y^{2}\}\nu(dy)<+\infty.\label{inffinvar}
\end{equation}
In our model, we assume that, under a certain risk-neutral measure
$\mathbb{Q}$, chosen by the market, the dynamics of the underlying
process $S=\left\{ S_{t}\right\} _{t\geq0}$ can be expressed as
\begin{equation}
S_{t}=e^{Y_{t}},\quad Y_{t}=\int_{0}^{t}\left(r_{s}-q\right)ds+X_{t},
\end{equation}
where $X=\left\{ X_{t}\right\} _{t\geq0}$ is a certain Lévy process,
$r=\left\{ r_{t}\right\} _{t\geq0}$ is the stochastic interest rate
process and $q$ the dividend yield. Then, one requires $\mathbb{E}\left[e^{X_{t}}\right]<+\infty$,
and, therefore, $\psi$ must admit analytic continuation into a strip
$\Im\xi\in(-1,0)$ and continuous continuation into the closed strip
$\Im\xi\in[-1,0]$.

The infinitesimal generator of $X$, denoted by $L$, is an integro-differential
operator that acts as follows:
\begin{equation}
Lu(x)=\frac{\sigma^{2}}{2}\frac{\partial^{2}u}{\partial x^{2}}(x)+\mu\frac{\partial u}{\partial x}(x)+\int_{-\infty}^{+\infty}\left(u(x+y)-u(x)-y\mathbf{1}_{|y|\leq1}\frac{\partial u}{\partial x}(x)\right)\nu(dy).\label{actL}
\end{equation}
 The operator $L$ can also be expressed as a pseudo-differential
operator (PDO) with the symbol$-\psi(\xi)$,\emph{ }that is $L=-\psi(D)$,
where $D=-i\partial_{x}$. Recall that a PDO $A=a(D)$ acts as follows:
\begin{equation}
A\left[u(x)\right]=(2\pi)^{-1}\int_{-\infty}^{+\infty}e^{ix\xi}a(\xi)\hat{u}(\xi)d\xi,\label{actPDO}
\end{equation}
 where $\hat{u}$ is the Fourier transform of a function $u$, that
is
\[
\hat{u}(\xi)=\int_{-\infty}^{+\infty}e^{-ix\xi}u(x)dx.
\]
It should be noted that the inverse Fourier transform, as specified
in (\ref{actPDO}), is classically defined only when both the symbol
$a(\xi)$ and the function $\hat{u}(\xi)$ exhibit certain desirable
properties. For instances, where these conditions are not met, the
inverse Fourier transform is more generally defined through the principle
of duality. 

Moreover, if the underlying stock does not distribute dividends, the
discounted price process is required to be a martingale, which implies
that the condition $\psi(-i)=0$ must be satisfied. This condition
facilitates the expression of the drift term $\mu$ in terms of the
other parameters, characterizing the Lévy process, as illustrated
in the following equation:
\begin{equation}
\mu=-\frac{\sigma^{2}}{2}+\int_{-\infty}^{+\infty}\left(1-e^{y}+y\mathbf{1}_{|y|\leq1}\right)\nu(dy).\label{drift}
\end{equation}
Hence, the infinitesimal generator may be rewritten as follows:
\begin{equation}
Lu\left(x\right)=\frac{\sigma^{2}}{2}\frac{\partial^{2}u}{\partial x^{2}}\left(x\right)-\frac{\sigma^{2}}{2}\frac{\partial u}{\partial x}\left(x\right)+\int_{\mathbb{R}}\left[u\left(x+y\right)-u\left(x\right)-\left(e^{y}-1\right)\frac{\partial u}{\partial x}\left(x\right)\right]\nu\left(dy\right).\label{actL2}
\end{equation}
The Appendix \ref{LP} shows the values of the $\psi$-function for
the main Lévy processes considered in the remainder of the paper,
under the assumption (\ref{drift}).

The stochastic interest rate is another key component of the proposed
market model. The Hull-White model is a good choice for modelling
the short interest rate $r$, due to its flexibility in fitting the
current term structure of interest rates, and its capacity to be easily
implemented in a tree or lattice, making it practical for valuing
interest rate derivatives. Its no-arbitrage framework ensures consistency
with market prices, allowing for a realistic modeling of interest
rate dynamics over time. The dynamics of the process $r$, the short
interest rate, reads as follows:
\[
dr_{t}=k_{HW}\left(\theta_{t}-r_{t}\right)dt+\sigma_{HW}dZ_{t}^{r},\ r_{0}=\bar{r}_{0},
\]
with $k_{HW}>0$ the speed of mean reversion, $\sigma_{HW}>0$ the
short rate volatility, $\bar{r}_{0}$ the initial value and $Z^{r}=\left\{ Z_{t}^{r}\right\} _{t\geq0}$
a Brownian motion. Moreover, $\theta_{t}$ is a deterministic function
which is completely determined by the market values of the zero-coupon
bonds by calibration. Now, let us consider $P^{M}\left(0,T\right)$,
the market price of the zero-coupon bond at time $0$ for the maturity
$T$ and $f^{M}\left(0,T\right)$, the market instantaneous forward
interest rate which is then defined by
\[
f^{M}\left(0,T\right)=-\frac{\partial\ln P^{M}\left(0,T\right)}{\partial T}.
\]
In the specified model, the process $r$ can be decomposed into the
sum of a Va\v{s}i\v{c}ek process and a deterministic function, delineated
as follows:
\[
r_{t}=\sigma_{HW}R_{t}+\beta\left(t\right),
\]
with $R=\left\{ R_{t}\right\} _{t\geq0}$ the stochastic process defined
by 
\[
dR_{t}=-k_{HW}R_{t}dt+dZ_{t}^{r},\ R_{0}=0,
\]
and $\beta\left(t\right)$ the function
\[
\beta\left(t\right)=f^{M}\left(0,t\right)+\frac{\sigma_{HW}^{2}}{2k^{2}}\left(1-\exp\left(-k_{HW}t\right)\right)^{2},
\]
 so obtained through calibration on bond prices.

A scenario referred to as the \emph{flat curve} is considered noteworthy.
In this context, we postulate that the market price of a zero-coupon
bond is determined by $P^{M}\left(t,T\right)=e^{-r_{0}\left(T-t\right)}$
and the initial forward rate is uniformly $f^{M}\left(0,T\right)=r_{0}$.
Given these presuppositions, the formulas for the coefficients $\beta\left(t\right)$
and $\ensuremath{\theta_{t}}$ are delineated as: 
\[
\beta\left(t\right)=r_{0}+\frac{\sigma_{HW}^{2}}{2k_{HW}^{2}}\left(1-\exp\left(-k_{HW}t\right)\right)^{2},\quad\text{and}\quad\theta_{t}=r_{0}+\frac{\sigma_{HW}^{2}}{2k_{HW}^{2}}\left(1-\exp\left(-2k_{HW}t\right)\right).
\]
We point out that the flat curve assumption is adopted for the numerical
experiments carried out in the dedicated section of this manuscript,
as it simplifies the parameter set considered.

\section{Contract formulation and valuation}

In the following, we consider a particular type of variable annuities,
called the \emph{equity-linked variable annuities} (ELVAs), which
offer policyholders an investment option that combines the elements
of conventional annuities with the potential for higher returns through
equity market exposure. A key characteristic of ELVAs includes the
GMDB and GMAB guarantees, ensuring minimum benefits regardless of
market conditions. As explored in the seminal work by Kirkby and Aguilar
\cite{kirkby2023valuation}, the surrender option is an essential
aspect of ELVAs, which allows policyholders to completely withdraw
their policy before maturity under certain conditions. Hereinafter,
we provide a brief description of the main features of the ELVA contract
and refer the interested reader to the original work for further details.

\subsection{The contract formulation}

The ELVA contract includes a minimum accumulation clause (GMAB), together
with a benefit in the event of the insured's premature death (GMDB).
At the initial time $t=0$, the policyholder purchasing the contract
pays a premium $P$ to the insurer. This premium is fully invested
in a fund, whose value $F=\left\{ F_{t}\right\} _{t\geq0}$ is equal
to $P$ at contract inception and it is linked to an underlying asset
$S$ which evolves over time as a Lévy process. The contract has a
pre-determined duration of $M$ years, but it may be terminated early
in the event of the insured's death or early surrender.

When an anniversary is reached, say the $m-th$ anniversary for $m\in\left\{ 1,\dots,M\right\} $,
the following actions take place:
\begin{enumerate}
\item Updating the value of the fund and withdrawal of fees. 

At each anniversary, the value of the fund $F$ is changed proportionally
to the value of the underlying asset $S$ to which it is connected
and is decreased by the fees proportionally to the parameter $\alpha_{m}$.
Specifically:
\[
F_{m}=\left(1-\alpha_{m-1}\right)F_{m-1}\cdot S_{m}/S_{m-1}.
\]
Although fees only reduce the value of the fund on an anniversary,
as a kind of discrete dividend, since the value of the fund between
two anniversaries is irrelevant to the valuation of the contract,
it is possible to think of fees paid continuously, as a kind of dividend
yield. Specifically, one can replace $q$ with $\hat{q}=q-\log\left(1-\alpha_{m}\right)$
in the dynamics of $S$ with respect to the time interval $\left[m-1,m\right]$,
and set
\[
F_{m}=F_{m-1}\cdot S_{m}/S_{m-1}.
\]
Within the framework of this particular model, by assuming that the
initial stock price $S_{0}$ is equivalent to the initial futures
price $F_{0}$, one obtains $F=S$, and, in particular, $F_{m}=S_{m}$
for all anniversaries dates $m$. This equivalence considerably simplifies
the notation, a convention that will be maintained throughout the
subsequent discourse. 
\item Payment of the death benefit in the event of the death of the policyholder.

If the death of the insured occurs between anniversaries $m-1$ and
$m$, the insured's heirs benefit from a payment $DB_{m}$, paid at
time $m$ and calculated as follows: 
\[
DB_{m}\left(F_{m}\right)=\max\left\{ F_{0}e^{gm},\min\left\{ F_{0}e^{cm},F_{m}\right\} \right\} .
\]
Then the contract ends. Please, note that $g$ and $c$ are minimal
and maximal growth rates respectively, so that the amount paid is
equal to the value of the fund, limited between two exponentially
increasing values.
\item Possible surrender (if $m<M$) or contract end (if $m=M$).

On each anniversary before maturity, the alive policyholder may decide
to terminate the contract early. In this case, the policyholder receives
a surrender benefit equal to the minimum between the value of the
fund and the maximum growth factor, reduced by a penalty governed
by the parameter $\gamma_{m}$. When the contract matures, payment
of the final payoff is automatic and the contract ends. In this case,
the surrender benefit is equal to death benefit. Therefore, $SB_{m}$,
the surrender benefit at time $m$, reads out:
\end{enumerate}
\[
SB_{m}\left(F_{m}\right)=\begin{cases}
\left(1-\gamma_{m}\right)\min\left\{ F_{0}e^{cm},F_{m}\right\}  & \text{if }m<M\\
DB_{m}\left(F_{m}\right) & \text{if }m=M.
\end{cases}
\]

\subsection{Modeling mortality}

Let $\delta$ represent the random time of death of the policyholder,
which is alive and $\omega$-years old at contract inception. Following
Kirkby and Aguilar \cite{kirkby2023valuation}, such a random variable
is assumed to follows a general probability mass function for an individual
of age $\omega$ at the time of policy purchase. This distribution,
represented as $\left\{ p_{m}^{\omega}:m\geq1\right\} $, captures
the probability of death occurring within the time interval $\mathcal{I}_{m}=\left(\omega+(m-1),\omega+m\right]$.
The model is designed to be neutral under both the risk-neutral measure
$\mathbb{Q}$ and the physical measure $\mathbb{P}$, enabling the
evaluation of mortality risk in variable annuity contracts. Specifically,
we assume that mortality risk is diversifiable. This approach provides
a realistic and flexible framework for incorporating mortality risk
into the pricing and valuation of life insurance products and annuity
contracts. 

\subsection{Early surrender and contract valuation}

The contract valuation strongly depends on modeling the surrender
strategy. Let $V\left(m,F_{m},r_{m}\right)$ denote the value of the
contract which has not yet been terminated at the $m-th$ anniversary,
so the policyholder is still alive and it has not surrendered yet.
Such a value is derived through a dynamic programming approach. The
optimal surrender strategy is defined as the stopping time $\tau^{S}$
that maximizes the expected discounted value of the surrender benefit
plus any mortality-related benefits, mathematically represented as:
\begin{equation}
V\left(0,F_{0},r_{0}\right)=\sup_{\tau\in\mathcal{T}}\mathbb{E}^{\mathbb{Q}}\left[e^{-\int_{0}^{\tau}r_{t}dt}SB_{\tau}\left(F_{\tau}\right)\ind_{\tau<\min\left(\delta,M\right)}+e^{-\int_{0}^{\min\left(\delta,M\right)}r_{t}dt}DB_{\min\left(\delta,M\right)}\left(F_{\min\left(\delta,M\right)}\right)\ind_{\tau\geq\min\left(\delta,M\right)}\right],\label{eq:AM}
\end{equation}
where $\mathcal{T}$ denotes the set of permissible surrender times
in $\left\{ 1,\dots,M\right\} $ and $\ind_{\left(\cdot\right)}$
is the indicator function. The optimal surrender strategy effectively
balances the potential economic gains from early surrender against
the contractual benefits of holding the policy until maturity or the
policyholder's death. 

It is worth noting that if one sets $\gamma_{m}=1$ for all value
of $m$, then the surrender benefit fades and therefore it is never
convenient to exercise in advance as one would be giving up the death
benefit, which is always positive. Therefore, to obtain a ``no-surrender''
version of the product, it is sufficient to set $\gamma_{m}=1$ for
all values of $m$. In such a case, we obtain

\begin{equation}
V\left(0,F_{0},r_{0}\right)=\mathbb{E}^{\mathbb{Q}}\left[e^{-\int_{0}^{\min\left(\delta,M\right)}r_{t}dt}DB_{\min\left(\delta,M\right)}\left(F_{\min\left(\delta,M\right)}\right)\right].\label{eq:EU}
\end{equation}

\section{The pricing algorithms}

We describe the two algorithms used to evaluate the contract discussed
in the previous Section, under the assumption that the underlying
fund $F$ follows a Lévy model between two anniversaries and the instantaneous
interest rate $r$ follows the Hull-White model.

\subsection{The Hybrid method}

The Hybrid method for pricing American option under a Lévy process
with stochastic interest rate is introduced by Briani et al. \cite{briani2019numerical},
where weak convergence and stability of the algorithm are also demonstrated.
Specifically, the method is proposed for the Bates-Hull-White model.
It is important to note that the Bates model assumes that the underlying
evolves as a Merton Jump Diffusion process with stochastic volatility
modelled by a Heston process, and the Hybrid method is stitched around
these assumptions. However, discarding the stochastic volatility assumption,
it can easily be adapted to consider any Lévy process for the underlying. 

In the original manuscript, the Hybrid method is based on the development
of a hybrid tree/finite difference approach and it employs a binomial
trees for discretize the stochastic volatility process and the stochastic
interest rate, combined with a space-continuous approximation for
the asset price process. Since we do not consider stochastic volatility
here, we only use a tree for the interest rate.

Let us define $\left\{ \hat{p}_{m}^{\omega}:m\geq1\right\} $ as the
conditional death probabilities, that is $\hat{p}_{m}^{\omega}$ is
the probability death occurring within the time interval $\left(\omega+(m-1),\omega+m\right]$
if it has not occurred yet:
\[
\hat{p}_{m}^{\omega}=\mathbb{P}\left(\omega+(m-1)<\delta\leq\omega+m\mid\delta\geq\omega+(m-1)\right).
\]
These probabilities can be easily calculated using the following recursive
relation:
\[
\hat{p}_{m}^{\omega}=\frac{p_{m}^{\omega}}{1-p_{m-1}^{\omega}}.
\]
First of all, we assume that fees are paid continuously, so that their
effect can be incorporated into the dynamics of the process $F$.
We observe that, by following risk-neutral valuation principles, during
the time between two anniversaries, the contract value $\mathcal{V}$
evolves as the price of a European option in the Lévy-Hull-White model.
Thus, following Briani et al. \cite{briani2019numerical}, $\mathcal{V}$
can be computed as the solution of the following problem:
\begin{equation}
\begin{cases}
\frac{\partial v}{\partial t}\left(t,Y,r\right) & +Lv+\left(r-q+\log\left(1-\alpha_{m}\right)\right)\frac{\partial v}{\partial y}+\frac{\sigma_{HW}^{2}}{2}r^{2}\frac{\partial^{2}v}{\partial r^{2}}+k_{HW}\left(\theta_{t}-r\right)\frac{\partial v}{\partial r}-rv=0\\
v\left(m+1,Y,r\right) & =\mathcal{V}\left(m+1,e^{Y},r\right),\\
\mathcal{V}\left(m,F_{m},r_{m}\right) & =\hat{p}_{m}^{\omega}\cdot DB_{m}\left(F_{m}\right)+\left(1-\hat{p}_{m}^{\omega}\right)\max\left(SB_{m}\left(F_{m}\right),v\left(m,\ln\left(F_{m}\right),r_{m}\right)\right),
\end{cases}\label{eq:4.1}
\end{equation}
with $Lu$ defined as in (\ref{actL2}). 

The first PIDE in (\ref{eq:4.1}) can be solved by means of the Hybrid
algorithm. This algorithm is based on the alternating spread of the
underlying and the interest rate. The former is handled by solving
a one-dimensional PIDE in which the interest rate is frozen, while
the latter is handled by using a bivariate tree.

The construction of the interest rate tree proceeds by discretizing
the time interval between two anniversaries into sub-intervals of
fixed length $\Delta t$, so that $N_{T}=1/\Delta t$ is the number
of time-steps per year. Here, $\Delta t$ is assumed to exactly divide
the time unity, so one can define a unique tree for all the duration
of the contract. In addition, by doing so, anniversary dates are included
in the tree structure. For each time step, the tree branches into
multiple possible future states, reflecting the possible movements
of the interest rate. The rate process $r$ at each node of the tree
is updated based on the discretized form of the Hull-White model's
stochastic differential equation, ensuring that the tree is recombining,
which means that it converges back to fewer states over time to maintain
computational efficiency. Specifically, the Hybrid algorithm exploits
the \textquotedblleft multiple-jumps\textquotedblright{} tree introduced
in Nelson and Ramaswamy \cite{nelson1990simple}: for $n=0,1,\ldots,M\cdot N_{T}$,
consider the lattice for the process $R$ defined by 
\[
\begin{array}{ll}
\mathcal{R}_{n}=\left\{ R_{j}^{n}\right\} _{j=0,1,\ldots,n} & \text{ with }R_{j}^{n}=(2j-n)\sqrt{\Delta t}\end{array}.
\]
For each fixed $R_{j}^{n}\in\mathcal{R}_{n}$, we denote the ``up''
and ``down'' jump by $R_{j_{u}(n,j)}^{n+1}$ and by $R_{j_{d}(n,j)}^{n+1}$,
the indices of which, the jump-indexes $j_{u}(n,j),j_{d}(n,j)$, are
defined as 
\[
\begin{aligned}j_{u}\left(n,j\right)= & \min\left\{ j^{*}:j+1\leq j^{*}\leq n+1\text{ and }R_{j}^{n}+\mu_{R}\left(R_{j}^{n}\right)\Delta t\leq R_{j^{*}}^{n+1}\right\} ,\\
j_{d}\left(n,j\right)= & \max\left\{ j^{*}:0\leq j^{*}\leq j\text{ and }R_{j}^{n}+\mu_{R}\left(R_{j}^{n}\right)\Delta t\geq R_{j^{*}}^{n+1}\right\} ,
\end{aligned}
\]
 where $\mu_{R}\left(R_{j}^{n}\right)=-k_{HW}R_{j}^{n}$ is the drift
of $R$, with the understanding $j_{u}(n,k)=n+1$ and respectively,
$j_{d}(n,k)=0$.

To ensure the tree accurately reflects the no-arbitrage condition
and aligns with the observed initial term structure of interest rates,
the transition probabilities between nodes are carefully calibrated.
This involves adjusting the probabilities such that the expected value
of the tree's future rates matches the market's forward rates, a process
that requires iterative adjustments to the model parameters. Starting
from node $(n,j)$, the probability that the process jumps to $j_{u}\left(n,j\right)$
and $j_{d}\left(n,j\right)$ at time-step $n+1$ are 
\[
p_{u}^{R}(n,j)=0\vee\frac{\mu_{R}\left(R_{j}^{n}\right)\Delta t+R_{j}^{n}-R_{j_{d}(n,j)}^{n+1}}{R_{j_{u}(n,j)}^{n+1}-R_{j_{d}(n,j)}^{n+1}}\wedge1\quad\text{ and }\quad p_{d}^{R}(n,j)=1-p_{u}^{R}(n,j).
\]
This interest rate tree, once constructed, integrates into the larger
framework of the Lévy models for option pricing. 

The diffusion of the underlying fund is handled by solving a local
PIDE written on the log price $Y=\log\left(F\right)$ which holds
between two time steps and it is obtained by assuming a constant interest
rate. In detail, consider a node $R_{j}^{n}$ and set $r_{j}^{n}=\sigma_{HW}R_{j}^{n}+\beta\left(n\Delta t\right)$.
Then, we solve 
\begin{multline}
\frac{\partial v}{\partial t}\left(t,Y,r_{j}^{n}\right)+\frac{\sigma^{2}}{2}\frac{\partial^{2}v}{\partial Y^{2}}+\left(r_{j}^{n}-q+\log\left(1-\alpha_{m}\right)-\frac{\sigma^{2}}{2}\right)\frac{\partial v}{\partial Y}\\
+\int_{\mathbb{R}}\left[v\left(t,Y+x,r_{j}^{n}\right)-v\left(t,Y,r_{j}^{n}\right)-\left(e^{x}-1\right)\frac{\partial v}{\partial Y}\right]\nu\left(dx\right)-r_{j}^{n}v=0.\label{eq:PIDE_v}
\end{multline}

\medskip

From an implementation point of view, we proceed by creating the tree
structure $\mathcal{R}$ and define a uniform grid $\mathcal{Y}=\left\{ y_{i},i=1,\dots,N_{Y}\right\} $
representing possible values for the process $Y$. In particular,
the grid $\mathcal{Y}$ is defined by following the localization step
described in \cite{cont2005finite} and we term $dy$ the space step
of such a grid. Let $\tilde{v}_{j,i}^{n}$ represent the numerical
approximation of the function $v$ at the node $\left(R_{j}^{n},y_{i}\right)$
at time $n\Delta t$.

The initial price is computed by moving backward in time, alternating
the interest rate diffusion and the solution of the PIDE (\ref{eq:PIDE_v}).
In particular, we start at the last time-step $M\cdot N_{T}$ by assigning
to each element $\left(R_{j}^{MN_{T}},y_{i}\right)$ of the product
$\left\{ R_{j}^{MN_{T}},j=0,\dots,N_{T}\right\} \times\mathcal{Y}$
the value of the final payoff, defined as equal to the death benefit
$DB_{M}\left(e^{y_{i}}\right)$, that is
\[
\tilde{v}_{j,i}^{MN_{T}}=DB_{M}\left(e^{y_{i}}\right).
\]
Then, we proceed to treat the nodes at second-to-last time-step, that
is $n=MN_{T}-1$. For each node $\left(R_{j}^{n},y_{i}\right)$, the
interest rate is diffused by calculating the weighted average of the
values at future nodes $\left(R_{j_{u}(n,j)}^{n+1},y_{i}\right)$
and $\left(R_{j_{d}(n,j)}^{n+1},y_{i}\right)$, weighted according
to the transition probabilities $p_{u}^{R}(n,j)$ and $p_{d}^{R}(n,j)$.
Then, for each node $R_{j}^{n}$ of the tree at time-step $n$ the
contract values at points $R_{j}^{n}\times\mathcal{Y}$ are updated
by solving the PIDE (\ref{eq:PIDE_v}). This is done by applying a
single step of the IMEX scheme proposed by Cont and Voltchkova \cite{cont2005finite}.
Specifically, such a scheme divides the differential equation into
stiff and non-stiff components: the stiff parts are treated implicitly,
while the non-stiff parts are handled explicitly. Following the approach
of Cont and Voltchkova, the boundary conditions chosen for the algorithm
are Neumann conditions. In detail, the value of the contract for extreme
values of $\mathcal{Y}$ are set equal to the value of the payoff
at maturity at those nodes, discounted to the risk-free rate by spreading
the interest rate through the binomial $\mathcal{R}$-tree: 
\begin{align*}
\tilde{v}_{j,1}^{n} & =e^{-r_{j}^{n}\Delta t}\left(p_{u}^{R}(n,j)\tilde{v}_{j_{u}(n,j),1}^{n}+p_{u}^{R}(d,j)\tilde{v}_{j_{d}(n,j),1}^{n}\right),\\
\tilde{v}_{j,N_{Y}}^{n} & =e^{-r_{j}^{n}\Delta t}\left(p_{u}^{R}(n,j)\tilde{v}_{j_{u}(n,j),N_{Y}}^{n}+p_{u}^{R}(d,j)\tilde{v}_{j_{d}(n,j),N_{Y}}^{n}\right).
\end{align*}
It is worth noting that the value of the contract outside $\mathcal{Y}$
is assumed to be constant and equal the closer boundary value, namely
$\tilde{v}_{j,1}^{n}$ and $\tilde{v}_{j,N_{Y}}^{n}$. This assumption
is necessary to handle the non-local nature of the Lévy operator $L$.

From the resolution of the PIDE, we obtain the value of the function
$\tilde{v}$ at time-step $MN_{T}-1$. By iterating the procedure
just described, from $n=MN_{T}-1,\dots,\left(M-1\right)N_{T}$ it
is possible to obtain the value of the function $\tilde{v}^{\left(M-1\right)N_{T}}$
at all points in $\left\{ R_{j}^{\left(M-1\right)N_{T}},j=0,\dots,\left(M-1\right)N_{T}\right\} \times\mathcal{Y}$.
Then, $\tilde{\mathcal{V}}_{j,i}^{\left(M-1\right)N_{T}}$, the numerical
approximation of the contract value $\mathcal{V}$ at the node $\left(R_{j}^{n},y_{i}\right)$
at time $M-1$, is computed as: 
\begin{equation}
\tilde{\mathcal{V}}_{j,i}^{\left(M-1\right)N_{T}}=\hat{p}_{M-1}^{\omega}\cdot DB_{M-1}\left(e^{y_{i}}\right)+\left(1-\hat{p}_{M-1}^{\omega}\right)\max\left(SB_{M-2}\left(e^{y_{i}}\right),\tilde{v}_{j,i}^{\left(M-1\right)N_{T}}\right).\label{eq:4.3}
\end{equation}

We stress out that, in the previous Formula (\ref{eq:4.3}), the death
benefit multiplies $\hat{p}_{M-1}^{\omega}$, which is the conditional
probability for $\delta$ to be in the time interval $\mathcal{I}_{M-1}$,
because at time $M-1$ the death benefit is paid out to the heirs
of those policyholders that died in $\mathcal{I}_{M-1}$. The procedure
previously described is repeated iteratively backward in time, considering
as starting time a generic starting anniversary $m$ instead of maturity
$M$, and replacing the terminal values computed through the payoff
function with the values $\left\{ \tilde{\mathcal{V}}_{j,i}^{mN_{T}},j=0,\dots,mN_{T}\text{ and }i=1,\dots,N_{Y}\right\} $
computed at previous step. This routine, should be repeated for up
to $m=1$. To finally obtain the numerical estimate of the price at
epoch zero, that is for $m=0$, one has to proceed similarly to what
has been done so far, but avoid considering any death benefit paid
at, and the possibility of a total lapse at contact initiation. Specifically,
equation (\ref{eq:4.3}) is replaced by setting $\tilde{\mathcal{V}}_{0,i}^{0}=\tilde{v}_{0,i}^{0}$
for all value $i=1,\dots,N_{Y}$. The initial contact value is finally
computed by interpolating the observed values $\left\{ \tilde{\mathcal{V}}_{0,i}^{0},i=1,\dots,N_{Y}\right\} $
on the grid $\mathcal{Y}$ at the value $\log\left(F_{0}\right)$.

\subsection{The Longstaff-Schwartz Monte Carlo method}

The valuation of the contract by means of a Monte Carlo methodology
is relatively simple if one excludes the possibility of early exercise,
i.e. one follows the formula (\ref{eq:EU}). The crucial element is
to have a set of simulations for the underlying process and the rate
process. In the literature, there are several established methods
for simulating Lévy processes (see e.g. Kienitz and Wetterau \cite{kienitz2013financial}
or Cools and Nuyens \cite{cools2016monte}), while for the simulation
of the interest rate under the Hull-White model, we refer to the exact
method by Ostrovski \cite{ostrovski2013efficient}. 

The procedure involves simulating the processes $F$, $r$ and $I=\left\{ I_{t}\right\} _{t\geq0}$
with $I_{t}=\int_{0}^{t}r_{s}ds$. Specifically, we consider $N_{MC}$
random paths $\left\{ \left(F_{m}^{k},r_{m}^{k},I_{m}^{k}\right),m=0,\dots,M\text{ and }k=1,\dots,N_{MC}\right\} $
and we approximate the initial $\mathcal{V}\left(0,F_{0},r_{0}\right)$
contract price with $\hat{\mathcal{V}}_{0}\left(F_{0},r_{0}\right)$
defined as
\[
\hat{\mathcal{V}}_{0}\left(F_{0},r_{0}\right)=\frac{1}{N_{MC}}\sum_{k=1}^{N_{MC}}\left[\sum_{m=1}^{M-1}p_{m}^{\omega}e^{-I_{m}^{k}}DB_{m}\left(F_{m}^{k}\right)+\left(1-\sum_{m=1}^{M-1}p_{m}^{\omega}\right)e^{-I_{M}^{k}}DB_{M}\left(F_{M}^{k}\right)\right].
\]

When, on the other hand, the possibility of early surrender is considered,
an optimal control strategy must be derived. Assigned $\tau_{m}$
a stopping time in $\left\{ m,m+1,\dots,M\right\} $, we define the
continuation value (for an alive policyholder) at the $m$-th anniversary
as the expected value of discounted cashflows obtained through the
application of the stopping time $\tau_{m}$, that is the following
function
\[
\mathcal{C}_{m}\left(a,b\right)=\mathbb{E}^{\mathbb{Q}}\left[\sum_{h=m+1}^{\tau-1}p_{h-m}^{\omega+m}e^{-\int_{m}^{h}r_{s}ds}DB_{h}\left(F_{h}\right)+\left(1-\sum_{h=m+1}^{\tau_{m}-1}p_{h-m}^{\omega+m}\right)e^{-\int_{m}^{\tau_{m}}r_{t}dt}SB_{\tau_{m}}\left(F_{\tau_{m}}\right)\mid F_{m}=a,r=b\right].
\]

It is well known that the optimal strategy is obtained by exercising
surrender as soon as the exercise value (i.e. the surrender benefit)
exceeds the continuation value (the expected discounted value of the
future payoff):
\begin{equation}
\tau_{m}=\min\left\{ h\in\mathbb{N}\text{ s.t. }h>m\text{ and }SB_{h}\left(F_{h}\right)\geq\mathcal{C}_{h}\left(F_{h},r_{h}\right)\right\} \cup\left\{ M\right\} .\label{eq:tau}
\end{equation}
The determination of the continuation value is the crucial aspect.
To estimate it, we apply the well-known Longstaff-Schwartz approach
\cite{longstaff2001valuing}, which estimates the continuation value
backward in time by means of a polynomial regression of discounted
future cashflows, calculated by using the optimal strategy at future
anniversaries. Specifically, we aim to compute the optimal polynomial
in the least squares sense. This polynomial is intended to regress
the set of points
\[
\left\{ y_{k}=\sum_{h=m+1}^{\tau-1}p_{h-m}^{\omega+m}e^{-\left(I_{h}^{k}-I_{m}^{k}\right)}DB_{h}\left(F_{h}^{k}\right)+\left(1-\sum_{h=m+1}^{\tau_{m}-1}p_{h-m}^{\omega+m}\right)e^{-\left(I_{\tau_{m}}^{k}-I_{m}^{k}\right)}SB_{\tau_{m}}\left(F_{\tau_{m}}^{k}\right),\ k=1,\dots,N_{MC}\right\} 
\]
i.e. the simulated discounted future cash-flows, against the set of
points
\[
\left\{ \left(F_{m}^{k},r_{m}^{k}\right),\ k=1,\dots,N_{MC}\right\} .
\]
The computation of the least squares bivariate polynomial is straightforward
and computational efficient (see e.g. Dyn \cite{dyn2001multivariate}).
Anyway, a direct application of the least squares regression proves
to be ineffective in this context since the continuation value of
the variable annuity considered here has an S-shape when the price
of the underlying fund $F$ changes, whereas it has a more linear
appearance when the interest rate changes. It is therefore difficult
to find a polynomial that well approximates such a shape. So, the
optimal stop strategy that is generated from such data is underperforming
and the price that is obtained is significantly lower than the price
associated with an optimal strategy.

Instead of addressing the problem of regression as a global problem
we propose to use a local regression approach: specifically, the domain
is partitioned into several sectors according to the value of the
underlying fund $F$ and a different polynomial in each sector is
estimated by using the least squares procedure. A similar technique
is used in other contexts by Hainaut and Akbaraly \cite{hainaut2023risk}.
The domain is divided solely according to the value of the underlying
fund considering the following threshold values, based on the minimum
and maximum growth rates: 
\[
b_{1}/8,b_{1}/4,b_{1}/2,b_{1},\frac{b_{1}+b_{2}}{2},b_{2},2b_{2},4b_{2},8b_{2},\quad\text{with \ensuremath{b_{1}}=\ensuremath{F_{0}e^{gm}\text{ and }b_{2}}=\ensuremath{F_{0}e^{cm}}}.
\]
Furthermore, if a sector contains more than $20\%$ of the total number
of points, then we divide it, determining the threshold value so that
half of the points are in the first sub-sector and half in the second.
The division of the domain is quite arbitrary, but effective. It responds
to the intention of capturing the various moneyness situations (deep
out of money, out of money, at the money, in the money, deep in the
money) with respect to both bounds, as well as avoiding the excessive
concentration of too many points in the same sector. In each sector
the regression is carried out using a complete two-variable polynomial.
Following a common approach in the Machine Learning field, the degree
of the polynomial exploited for regression in each sector is determined
according a $80\%-20\%$ criterion. This means that the sample data
is randomly divided in two sets: $80\%$ of the data is used for regression,
$20\%$ for out-of-sample testing. Starting from a polynomial of degree
zero, the degree is increased as long as the mean squared error on
the sample group continues to decrease. In this way, the correct polynomial
degree is selected with respect to the predictive capacity of the
polynomial and overfitting is avoided. Once the best polynomial degree
is determined, all the $100\%$ of the data-set is used for the regression.

Once an effective strategy for estimating the continuation value is
available, by proceeding backward in time from $m=M$ to $m=1$, one
can approximate the optimal stop strategy $\tau_{0}$ as in (\ref{eq:tau})
and consequently the initial contract price as 
\[
\hat{\mathcal{V}}_{0}\left(F_{0},r_{0}\right)=\frac{1}{N_{MC}}\sum_{k=1}^{N_{MC}}\left[\sum_{h=1}^{\tau-1}p_{h}^{\omega}e^{-\left(I_{h}^{k}-I_{0}^{k}\right)}DB_{h}\left(F_{h}^{k}\right)+\left(1-\sum_{h=1}^{\tau_{0}-1}p_{h}^{\omega}\right)e^{-\left(I_{\tau_{0}}^{k}-I_{0}^{k}\right)}SB_{\tau_{0}}\left(F_{\tau_{0}}^{k}\right)\right].
\]

\section{Numerical Results}

In this Section, we focus on the calculation of the surrender premium,
that is the difference in price between the contract under optimal
surrender, and the contract that does not allow early surrender. The
surrender premium plays a significant role in the pricing and design
of ELVA contracts: by assessing the incentive for policyholders to
surrender their policies early, insurers can adjust premiums, fees,
and benefits to make the contracts more attractive while maintaining
profitability. Understanding the surrender premium is essential for
designing policies that align with policyholders' interests. It can
be used to structure contracts that disincentivize early surrender,
thus encouraging long-term investment, which is beneficial for both
the insurer and the policyholder. Regulators require insurers to maintain
adequate reserves for their obligations. Accurately calculating the
surrender premium is part of ensuring that these reserves are sufficient
to cover potential surrenders, thereby complying with regulatory requirements.
In the competitive market of insurance products, offering ELVAs with
favorable surrender terms can be a differentiator. Accurate calculation
of the surrender premium allows insurers to design products that are
competitive and appealing to potential customers. 

First of all, we calculate the surrender premium, that is the difference
between the ``surrender'' price (optimal surrender available, for
$\gamma_{m}=0.02$ for all values of $m$) and the ``no-surrender''
price of the contract (surrender is not allowed, or is never convenient,
that is $\gamma_{m}=1$ for all values of $m$). Then, we focus on
a comparative statistics analysis to understand how the surrender
premium varies as individual market parameters vary.

In the first analysis, we focus on four Lévy models, while in the
second analysis, for the sake of simplicity, we consider only the
NIG Lévy model. The parameters of these stochastic models are the
same as those employed by Kirkby in other analyses of Levy's models\footnote{For further details, see Kirkby's repository at https://github.com/jkirkby3/PROJ\_Option\_Pricing\_Matlab.},
while the parameters of the Hull-White model were obtained through
a standard calibration procedure\footnote{See https://it.mathworks.com/help/fininst/calibrating-hull-white-model-using-market-data.html
for further details.} from the prices of caplets on Euro (Data downloaded on 02-Feb-2024)
and are thus plausible values for real applications, with the exception
of $r_{0}$ that was chosen equal to $2\%$ as in Kirkby and Aguilar
\cite{kirkby2023valuation}. As for the zero-coupon bond price curve,
we choose to use a flat curve in order to facilitate replication of
our results.

The algorithms have been implemented in MATLAB and computations have
been preformed on a server which employs a $2.40$ GHz Intel Xenon
processor (Gold 6148, Skylake) and 64 GB of RAM. The mortality Table\footnote{The mortality Table can be accessed at https://github.com/jkirkby3/PROJ\_Option\_Pricing\_Matlab
.} employed herein aligns with that used by Kirkby and Aguilar \cite{kirkby2023valuation}.

\subsection{Computation of the surrender premium }

We advance to assess the surrender benefits within four distinct Lévy
models alongside the Hull-White model: namely, the Normal Inverse
Gaussian (NIG), Variance Gamma (VG), CGMY, and Merton Jump Diffusion
(MJD) models. These evaluations are performed by using both the Hybrid
method and the Longstaff-Schwartz Monte Carlo (LSMC) technique, across
five unique numerical settings tailored for the pricing algorithms.
The initial four settings, labeled A through D, are designed to achieve
specific computation run times (4s, 15s, 60s, and 240s, respectively)
for both methodologies when applied to the NIG model, with parameters
set at $c=15\%$ and $g=1\%$. The fifth setting is free from time
constraints, serving as a benchmark for comparison. This methodical
selection of diverse configurations allows for an in-depth evaluation
of the numerical efficiency of each algorithm under review. Table
\ref{tab:0} outlines the parameters for the numerical algorithms.
It is noteworthy that for configurations A through D, identical simulations
were utilized to ascertain both the exercise strategy and the pricing.
For the benchmark, however, $4.0\times10^{7}$ trajectories were simulated
to identify the optimal exercise strategy, and $1.0\times10^{7}$
out-of-sample trajectories were used for price calculation to mitigate
potential benchmark distortion from, even minimal, overfitting. Given
the extensive simulation times for the CGMY model, the benchmark simulation
count is reduced by a factor of 4; despite this adjustment, the benchmark
computation still spans several days. 

Tables \ref{tab:NIG}, \ref{tab:VG}, \ref{tab:CGMY} and \ref{tab:MJD}
report the model parameters, as well as the estimated values of the
surrender premium and, for the Longstaff-Schwartz Monte Carlo method,
also the $99\%$ confidence intervals. The data shows that the Hybrid
benchmark always falls within the confidence interval of the Longstaff-Schwartz
Monte Carlo benchmark, so the two benchmarks agree. 

The Hybrid method outperforms the Monte Carlo method: in general,
in the 4 configurations considered, the values returned by the hybrid
method are closer to the benchmark and monotonic convergence is highlighted,
unlike the Monte Carlo method. Furthermore, we observe that, if we
consider the Hybrid method, already configuration B provides accurate
results, so we use it for further testing.

\begin{table}
\begin{centering}
\begin{tabular}{cccc}
\toprule 
Configuration & Target time & LSMC & Hybrid\tabularnewline
\midrule
A & $4s$ & $4.3\cdot10^{4}$ & $0.015,\phantom{11}7$\tabularnewline
B & $15s$ & $2.5\cdot10^{5}$ & $0.010,\phantom{1}10$\tabularnewline
C & $60s$ & $6.7\cdot10^{5}$ & $0.008,\phantom{1}15$\tabularnewline
D & $240s$ & $2.0\cdot10^{6}$ & $0.005,\phantom{1}22$\tabularnewline
\midrule
Benchmarks &  & $4.0\cdot10^{7}$ & $0.001,100$\tabularnewline
\bottomrule
\end{tabular}
\par\end{centering}
\caption{\label{tab:0}parameter for the numerical algorithms. For each configuration,
the table reports: the target computational time for the computation
of the surrender premium; the number of simulations for the Monte
Carlo algorithm; the space increment $dy$ and the number of time
steps per period $N_{T}$.}
\end{table}

\begin{table}
\begin{centering}
\resizebox{\textwidth}{!}{%
\begin{tabular}{ccccccccc}
\toprule 
 &  & \multicolumn{3}{c}{$g=1\%$} &  & \multicolumn{3}{c}{$g=3\%$}\tabularnewline
\cmidrule{3-5} \cmidrule{4-5} \cmidrule{5-5} \cmidrule{7-9} \cmidrule{8-9} \cmidrule{9-9} 
 &  & LSMC & Hybrid & Benchmarks &  & LSMC & Hybrid & Benchmarks\tabularnewline
$c=\phantom{0}5\%$ &  &  &  & {\scriptsize{}(Hybrid, LSMC)} &  &  &  & {\scriptsize{}(Hybrid, LSMC)}\tabularnewline
\midrule
A &  & $0.1521\pm0.0048$ & $0.1526$ &  &  & $0.0449\pm0.0028$ & $0.0458$ & \tabularnewline
B &  & $0.1527\pm0.0020$ & $0.1523$ & $0.1520$ &  & $0.0452\pm0.0012$ & $0.0457$ & $0.0454$\tabularnewline
C &  & $0.1523\pm0.0012$ & $0.1523$ & $0.1520\pm0.0003$ &  & $0.0451\pm0.0007$ & $0.0456$ & $0.0454\pm0.0002$\tabularnewline
D &  & $0.1519\pm0.0007$ & $0.1522$ &  &  & $0.0456\pm0.0004$ & $0.0455$ & \tabularnewline
\midrule
 &  &  &  &  &  &  &  & \tabularnewline
$c=15\%$ &  &  &  &  &  &  &  & \tabularnewline
\midrule
A &  & $0.1961\pm0.0170$ & $0.1888$ &  &  & $0.1390\pm0.0153$ & $0.1305$ & \tabularnewline
B &  & $0.1893\pm0.0075$ & $0.1888$ & $0.1887$ &  & $0.1277\pm0.0069$ & $0.1304$ & $0.1302$\tabularnewline
C &  & $0.1895\pm0.0046$ & $0.1888$ & $0.1889\pm0.0012$ &  & $0.1304\pm0.0043$ & $0.1304$ & $0.1301\pm0.0011$\tabularnewline
D &  & $0.1887\pm0.0027$ & $0.1888$ &  &  & $0.1307\pm0.0025$ & $0.1303$ & \tabularnewline
\midrule
 &  &  &  &  &  &  &  & \tabularnewline
$c=30\%$ &  &  &  &  &  &  &  & \tabularnewline
\midrule
A &  & $0.2007\pm0.0986$ & $0.1867$ &  &  & $0.1425\pm0.0965$ & $0.1497$ & \tabularnewline
B &  & $0.1911\pm0.0337$ & $0.1870$ & $0.1874$ &  & $0.1539\pm0.0342$ & $0.1499$ & $0.1502$\tabularnewline
C &  & $0.1875\pm0.0227$ & $0.1871$ & $0.1858\pm0.0058$ &  & $0.1509\pm0.0227$ & $0.1500$ & $0.1472\pm0.0059$\tabularnewline
D &  & $0.1866\pm0.0128$ & $0.1873$ &  &  & $0.1485\pm0.0126$ & $0.1501$ & \tabularnewline
\bottomrule
\end{tabular}}
\par\end{centering}
\caption{\label{tab:NIG}surrender premium computed for the NIG-HW model with
$\alpha_{NIG}=6$, $\beta_{NIG}=-0.4$, $\delta_{NIG}=2$, $r_{0}=0.02$,
$k_{HW}=0.2$, $\sigma_{HW}=0.03$, $q=0.01$, $M=25$, $\omega=30$,
$\alpha_{m}=0.02$, $\gamma_{m}=0.02$, $g=1\%,3\%$, $c=5\%,15\%,30\%$.}
\end{table}

\begin{table}
\begin{centering}
\resizebox{\textwidth}{!}{%
\begin{tabular}{ccccccccc}
\toprule 
 &  & \multicolumn{3}{c}{$g=1\%$} &  & \multicolumn{3}{c}{$g=3\%$}\tabularnewline
\cmidrule{3-5} \cmidrule{4-5} \cmidrule{5-5} \cmidrule{7-9} \cmidrule{8-9} \cmidrule{9-9} 
 &  & LSMC & Hybrid & Benchmarks &  & LSMC & Hybrid & Benchmarks\tabularnewline
$c=\phantom{0}5\%$ &  &  &  & {\scriptsize{}(Hybrid, LSMC)} &  &  &  & {\scriptsize{}(Hybrid, LSMC)}\tabularnewline
\midrule
A &  & $0.1299\pm0.0043$ & $0.1327$ &  &  & $0.0425\pm0.0025$ & $0.0445$ & \tabularnewline
B &  & $0.1331\pm0.0018$ & $0.1327$ & $0.1325$ &  & $0.0435\pm0.0010$ & $0.0444$ & $0.0441$\tabularnewline
C &  & $0.1330\pm0.0011$ & $0.1326$ & $0.1325\pm0.0003$ &  & $0.0441\pm0.0006$ & $0.0443$ & $0.0441\pm0.0002$\tabularnewline
D &  & $0.1325\pm0.0006$ & $0.1326$ &  &  & $0.0440\pm0.0004$ & $0.0442$ & \tabularnewline
\midrule
 &  &  &  &  &  &  &  & \tabularnewline
$c=15\%$ &  &  &  &  &  &  &  & \tabularnewline
\midrule
A &  & $0.1265\pm0.0065$ & $0.1303$ &  &  & $0.0553\pm0.0052$ & $0.0591$ & \tabularnewline
B &  & $0.1313\pm0.0027$ & $0.1304$ & $0.1307$ &  & $0.0588\pm0.0022$ & $0.0591$ & $0.0592$\tabularnewline
C &  & $0.1316\pm0.0016$ & $0.1305$ & $0.1309\pm0.0004$ &  & $0.0588\pm0.0013$ & $0.0592$ & $0.0591\pm0.0003$\tabularnewline
D &  & $0.1305\pm0.0010$ & $0.1306$ &  &  & $0.0592\pm0.0008$ & $0.0592$ & \tabularnewline
\midrule
 &  &  &  &  &  &  &  & \tabularnewline
$c=30\%$ &  &  &  &  &  &  &  & \tabularnewline
\midrule
A &  & $0.1357\pm0.0069$ & $0.1383$ &  &  & $0.0596\pm0.0056$ & $0.0649$ & \tabularnewline
B &  & $0.1392\pm0.0029$ & $0.1385$ & $0.1389$ &  & $0.0634\pm0.0023$ & $0.0650$ & $0.0652$\tabularnewline
C &  & $0.1399\pm0.0017$ & $0.1386$ & $0.1387\pm0.0005$ &  & $0.0644\pm0.0014$ & $0.0651$ & $0.0651\pm0.0004$\tabularnewline
D &  & $0.1386\pm0.0010$ & $0.1387$ &  &  & $0.0641\pm0.0008$ & $0.0651$ & \tabularnewline
\bottomrule
\end{tabular}}
\par\end{centering}
\caption{\label{tab:VG}surrender premium computed for the VG-HW model with
$\kappa_{VG}=0.85$ $\theta_{VG}=0$ $\sigma_{VG}=0.2$, $r_{0,HW}=0.02$,
$k_{HW}=0.2$, $\sigma_{HW}=0.03$, $\Delta=1$, $M=25$, $\omega=30$,
$\alpha_{m}=0.02$, $\gamma_{m}=0.02$, $g=1\%,3\%$, $c=5\%,15\%,30\%$.}
\end{table}

\begin{table}
\begin{centering}
\resizebox{\textwidth}{!}{%
\begin{tabular}{ccccccccc}
\toprule 
 &  & \multicolumn{3}{c}{$g=1\%$} &  & \multicolumn{3}{c}{$g=3\%$}\tabularnewline
\cmidrule{3-5} \cmidrule{4-5} \cmidrule{5-5} \cmidrule{7-9} \cmidrule{8-9} \cmidrule{9-9} 
 &  & LSMC & Hybrid & Benchmarks &  & LSMC & Hybrid & Benchmarks\tabularnewline
$c=\phantom{0}5\%$ &  &  &  & {\scriptsize{}(Hybrid, LSMC)} &  &  &  & {\scriptsize{}(Hybrid, LSMC)}\tabularnewline
\midrule
A &  & $0.1421\pm0.0039$ & $0.1410$ &  &  & $0.0351\pm0.0021$ & $0.0358$ & \tabularnewline
B &  & $0.1410\pm0.0017$ & $0.1411$ & $0.1413$ &  & $0.0355\pm0.0009$ & $0.0358$ & $0.0356$\tabularnewline
C &  & $0.1415\pm0.0010$ & $0.1411$ & $0.1413\pm0.0004$ &  & $0.0355\pm0.0005$ & $0.0357$ & $0.0356\pm0.0002$\tabularnewline
D &  & $0.1411\pm0.0006$ & $0.1411$ &  &  & $0.0356\pm0.0003$ & $0.0357$ & \tabularnewline
\midrule
 &  &  &  &  &  &  &  & \tabularnewline
$c=15\%$ &  &  &  &  &  &  &  & \tabularnewline
\midrule
A &  & $0.1432\pm0.0040$ & $0.1416$ &  &  & $0.0360\pm0.0022$ & $0.0369$ & \tabularnewline
B &  & $0.1422\pm0.0017$ & $0.1418$ & $0.1422$ &  & $0.0368\pm0.0009$ & $0.0368$ & $0.0366$\tabularnewline
C &  & $0.1425\pm0.0010$ & $0.1420$ & $0.1424\pm0.0004$ &  & $0.0365\pm0.0005$ & $0.0368$ & $0.0365\pm0.0002$\tabularnewline
D &  & $0.1422\pm0.0006$ & $0.1421$ &  &  & $0.0365\pm0.0003$ & $0.0367$ & \tabularnewline
\midrule
 &  &  &  &  &  &  &  & \tabularnewline
$c=30\%$ &  &  &  &  &  &  &  & \tabularnewline
\midrule
A &  & $0.1434\pm0.0040$ & $0.1423$ &  &  & $0.0354\pm0.0022$ & $0.0370$ & \tabularnewline
B &  & $0.1429\pm0.0017$ & $0.1425$ & $0.1428$ &  & $0.0370\pm0.0008$ & $0.0369$ & $0.0367$\tabularnewline
C &  & $0.1432\pm0.0010$ & $0.1426$ & $0.1428\pm0.0004$ &  & $0.0365\pm0.0005$ & $0.0368$ & $0.0367\pm0.0002$\tabularnewline
D &  & $0.1427\pm0.0006$ & $0.1427$ &  &  & $0.0366\pm0.0003$ & $0.0368$ & \tabularnewline
\bottomrule
\end{tabular}}
\par\end{centering}
\caption{\label{tab:CGMY}surrender premium computed for the CGMY-HW model
with $C_{CGMY}=0.02$, $G_{CGMY}=5$, $M_{CGMY}=15$, $Y_{CGMY}=1.2$,
$r_{0,HW}=0.02$, $k_{HW}=0.2$, $\sigma_{HW}=0.03$, $q=0.01$, $M=25$,
$\omega=30$, $\alpha_{m}=0.02$, $\gamma_{m}=0.02$, $g=1\%,3\%$,
$c=5\%,15\%,30\%$.}
\end{table}

\begin{table}
\begin{centering}
\resizebox{\textwidth}{!}{%
\begin{tabular}{ccccccccc}
\toprule 
 &  & \multicolumn{3}{c}{$g=1\%$} &  & \multicolumn{3}{c}{$g=3\%$}\tabularnewline
\cmidrule{3-5} \cmidrule{4-5} \cmidrule{5-5} \cmidrule{7-9} \cmidrule{8-9} \cmidrule{9-9} 
 &  & LSMC & Hybrid & Benchmarks &  & LSMC & Hybrid & Benchmarks\tabularnewline
$c=\phantom{0}5\%$ &  &  &  & {\scriptsize{}(Hybrid, LSMC)} &  &  &  & {\scriptsize{}(Hybrid, LSMC)}\tabularnewline
\midrule
A &  & $0.1387\pm0.0046$ & $0.1377$ &  &  & $0.0499\pm0.0028$ & $0.0490$ & \tabularnewline
B &  & $0.1366\pm0.0019$ & $0.1377$ & $0.1375$ &  & $0.0493\pm0.0011$ & $0.0489$ & $0.0488$\tabularnewline
C &  & $0.1370\pm0.0012$ & $0.1376$ & $0.1375\pm0.0003$ &  & $0.0487\pm0.0007$ & $0.0489$ & $0.0487\pm0.0002$\tabularnewline
D &  & $0.1371\pm0.0007$ & $0.1376$ &  &  & $0.0484\pm0.0004$ & $0.0488$ & \tabularnewline
\midrule
 &  &  &  &  &  &  &  & \tabularnewline
$c=15\%$ &  &  &  &  &  &  &  & \tabularnewline
\midrule
A &  & $0.1325\pm0.0098$ & $0.1293$ &  &  & $0.0715\pm0.0084$ & $0.0695$ & \tabularnewline
B &  & $0.1290\pm0.0041$ & $0.1294$ & $0.1298$ &  & $0.0699\pm0.0036$ & $0.0696$ & $0.0698$\tabularnewline
C &  & $0.1288\pm0.0025$ & $0.1296$ & $0.1298\pm0.0006$ &  & $0.0694\pm0.0022$ & $0.0697$ & $0.0699\pm0.0004$\tabularnewline
D &  & $0.1288\pm0.0015$ & $0.1297$ &  &  & $0.0689\pm0.0013$ & $0.0697$ & \tabularnewline
\midrule
 &  &  &  &  &  &  &  & \tabularnewline
$c=30\%$ &  &  &  &  &  &  &  & \tabularnewline
\midrule
A &  & $0.1480\pm0.0108$ & $0.1428$ &  &  & $0.0802\pm0.0097$ & $0.0809$ & \tabularnewline
B &  & $0.1428\pm0.0048$ & $0.1431$ & $0.1437$ &  & $0.0785\pm0.0043$ & $0.0811$ & $0.0813$\tabularnewline
C &  & $0.1431\pm0.0029$ & $0.1433$ & $0.1438\pm0.0008$ &  & $0.0798\pm0.0026$ & $0.0812$ & $0.0810\pm0.0007$\tabularnewline
D &  & $0.1429\pm0.0017$ & $0.1435$ &  &  & $0.0799\pm0.0015$ & $0.0812$ & \tabularnewline
\bottomrule
\end{tabular}}
\par\end{centering}
\caption{\label{tab:MJD}surrender premium computed for the MJD-HW model with
$\sigma_{MJD}=0.25$, $\lambda_{MJD}=0.6$, $\mu_{MJD}^{j}=0.01$,
$\sigma_{MJD}^{j}=0.13$, $r_{0,HW}=0.02$, $k_{HW}=0.2$, $\sigma_{HW}=0.03$,
$q=0.01$, $M=25$, $\omega=30$, $\alpha_{m}=0.02$, $\gamma_{m}=0.02$,
$g=1\%,3\%$, $c=5\%,15\%,30\%$.}
\end{table}

\subsection{Comparative statics analysis of interest rate parameters on the surrender
premium}

In this Section, we focus on the NIG model specifically within the
framework of numerical configuration B, as previously analyzed. In
particular, when not otherwise indicated, we employ the following
parameters: $\alpha_{NIG}=6$, $\beta_{NIG}=-0.4$, $\delta_{NIG}=2$,
$r_{0}=0.02$, $k_{HW}=0.2$, $\sigma_{HW}=0.03$, $q=0.01$, $M=25$,
$\omega=30$, $\alpha_{m}=0.02$, $\gamma_{m}=0.02$ and $c=15\%$.
Here, we undertake a graphical exploration to discern how the surrender
premium is influenced by varying some of the most interesting market
parameters.

\begin{figure}
\begin{centering}
\includegraphics[width=1\textwidth]{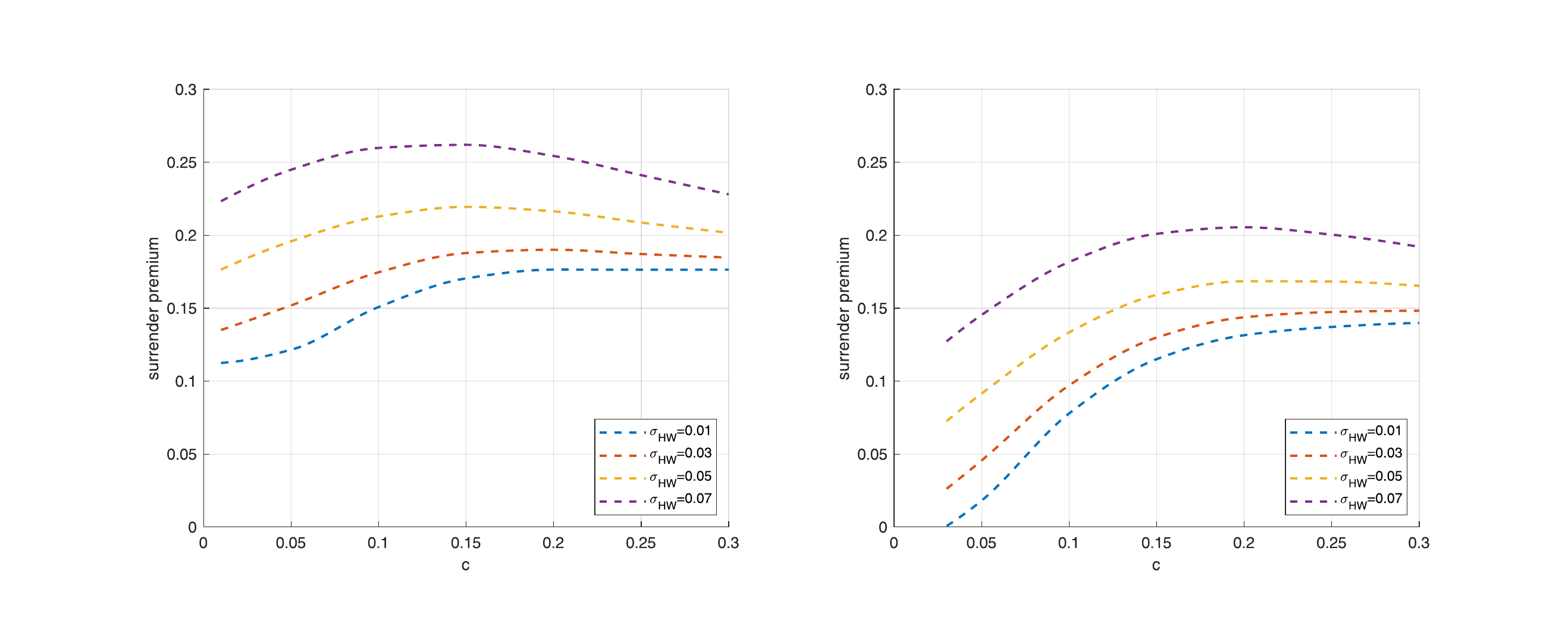}
\par\end{centering}
\begin{centering}
\includegraphics[width=1\textwidth]{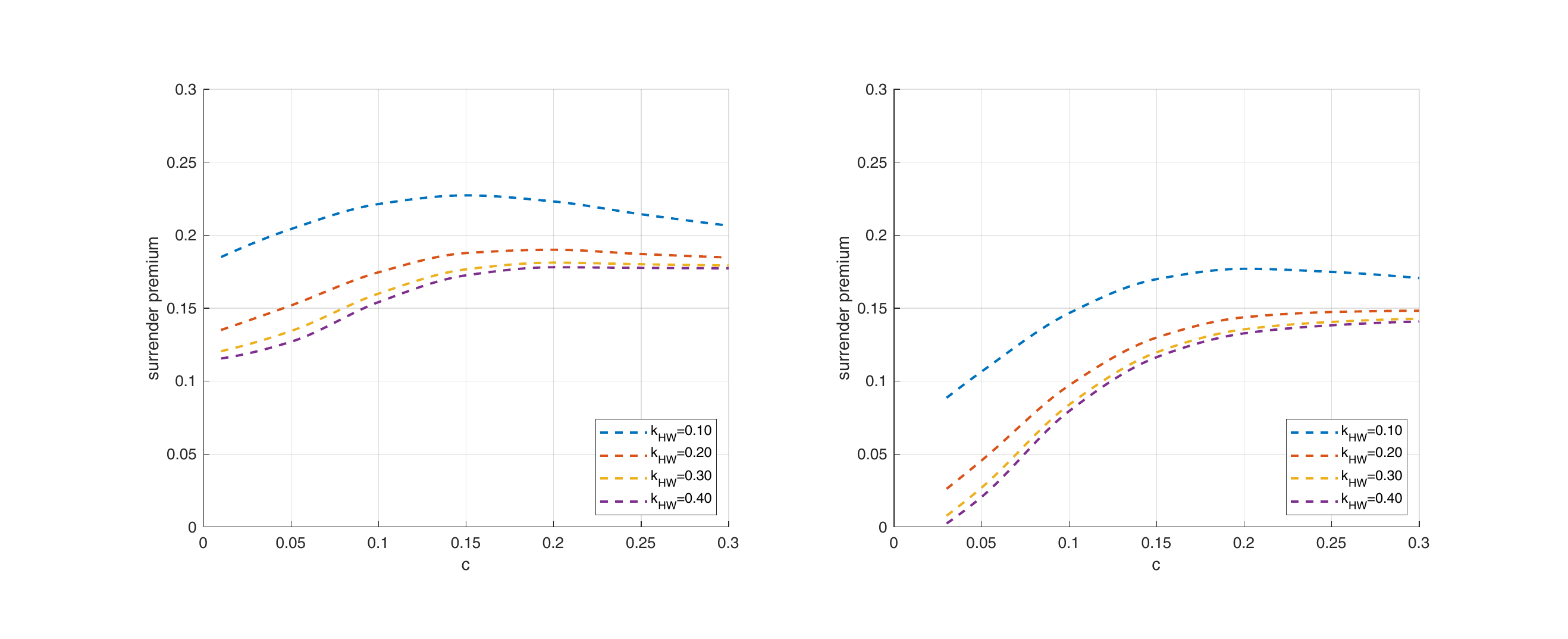}
\par\end{centering}
\caption{\label{fig_7}surrender premium as a function of $c$ while changing
$\sigma_{HW}$ (first row) and $k_{HW}$ (second row) for $g=1\%$
(first column) and $g=3\%$ (second column).}
\end{figure}

In Figure \ref{fig_7}, we investigate how the surrender benefit varies
for different values of $c$ as we change the value of the parameters
$\sigma_{HW}$ and $k_{HW}$. From these plots, we can observe that
the surrender benefit has a non-monotonic concave trend: in all the
cases analyzed, a maximum value is observed as the value of $c$ is
around $15\%-20\%$. Moreover, the surrender benefit increases as
$\sigma_{HW}$ increases and decreases as $k_{HW}$ increases: the
greater the volatility of the rate, the greater the incentive to surrender.
This phenomenon can be observed in practice: policyholders tend to
surrender when interest rates are high because the guarantees offered
by the insurance product are less useful in protecting the principal,
and these high values are more likely to happen for large values of
$\sigma_{HW}$ and $k_{HW}$. Therefore, insurers should be fearful
of situations in which interest rates change abruptly and consider
introducing some forms of premiums that incentivize customers not
to withdraw from the contract when it appears cheaper. 

\begin{figure}
\begin{centering}
\includegraphics[width=1\textwidth]{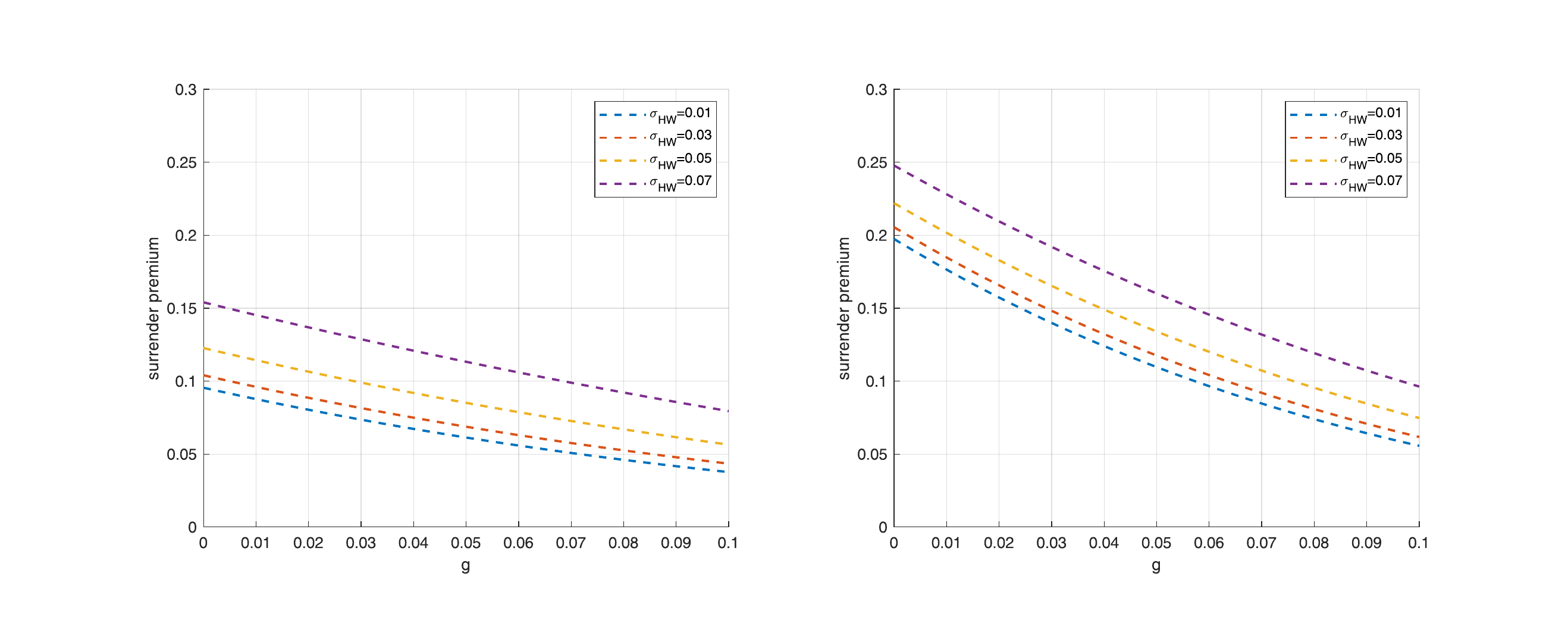}
\par\end{centering}
\begin{centering}
\includegraphics[width=1\textwidth]{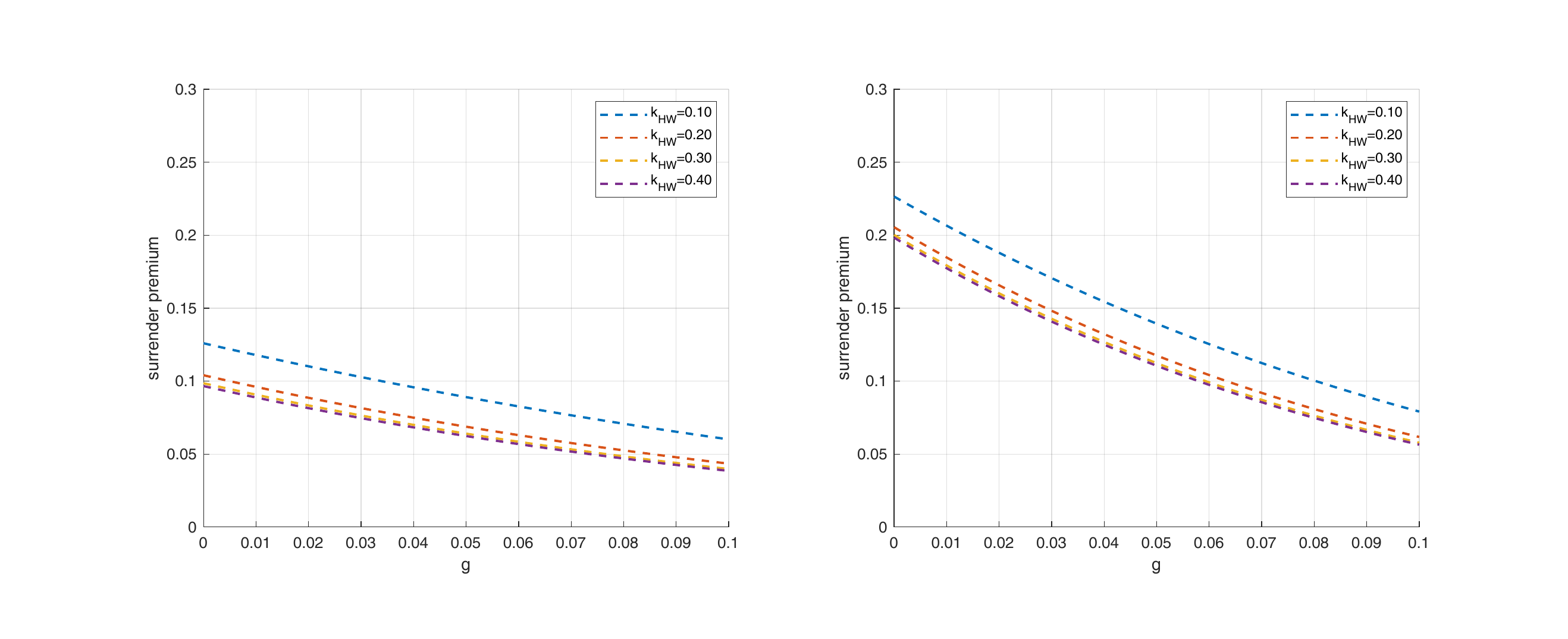}
\par\end{centering}
\caption{\label{fig_8}surrender premium as a function of $g$ while changing
$\sigma_{HW}$ (first row) and $k_{HW}$ (second row) for $\alpha_{m}=0\%$
(first column) and $\alpha_{m}=2\%$ (second column).}
\end{figure}

\begin{figure}
\begin{centering}
\includegraphics[width=1\textwidth]{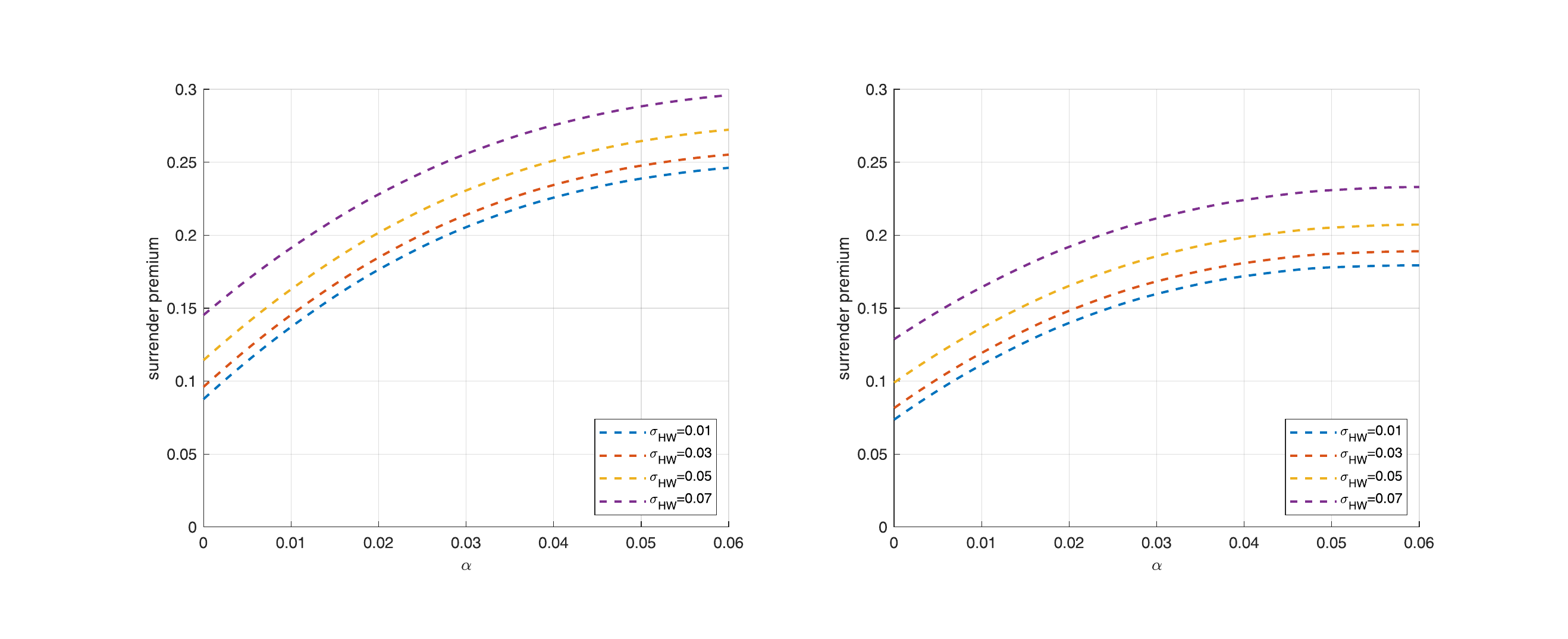}
\par\end{centering}
\begin{centering}
\includegraphics[width=1\textwidth]{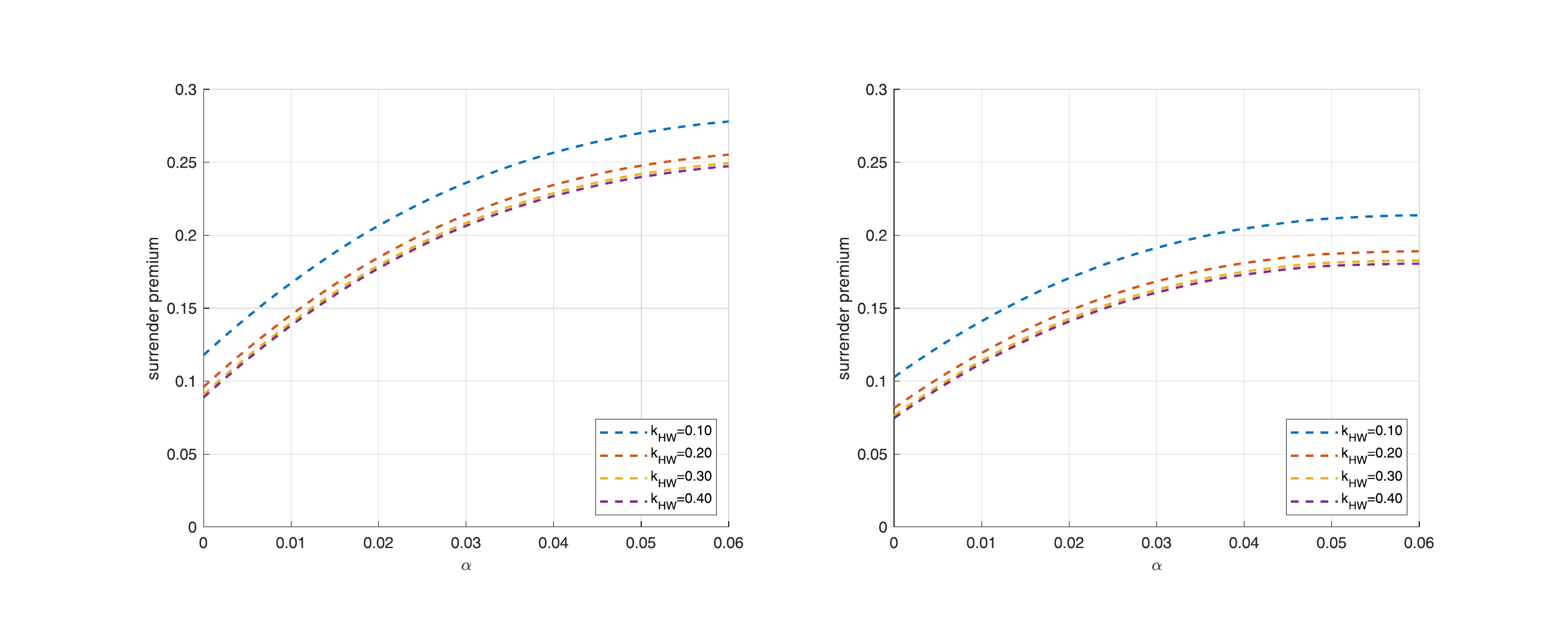}
\par\end{centering}
\caption{\label{fig_9}surrender premium as a function of $\alpha_{m}$ while
changing $\sigma_{HW}$ (first row) and $k_{HW}$ (second row) for
$g=1\%$ (first column) and $g=3\%$ (second column).}
\end{figure}

In Figure \ref{fig_8}, we observe the surrender benefit as a function
of the growth floor, $g$, for different settings of $\sigma_{HW}$,
$k_{HW}$ and $\alpha_{m}$. In all the plots, the surrender premium
decreases as $g$ varies, indicating that the greater the protection
on minimum capital growth, the lower the incentive to surrender. The
dependence of the surrender premium on the fees applied by the insurer
is highlighted in Figure \ref{fig_9}, where the surrender benefit
is plotted against the fee parameter $\alpha_{m}.$ In all images,
the surrender premium increases as a varies, indicating that the higher
the cost of the contract the greater the incentive to surrender.

Furthermore, we also studied the surrender premium as the $\alpha_{NIG}$
parameter varies, finding results similar to those produced by Kirkby
and Aguilar \cite{kirkby2023valuation} in the non-stochastic rate
Lévy model.

\begin{figure}
\begin{centering}
\subfloat[$m=5$.]{\begin{centering}
\includegraphics[width=0.35\textwidth]{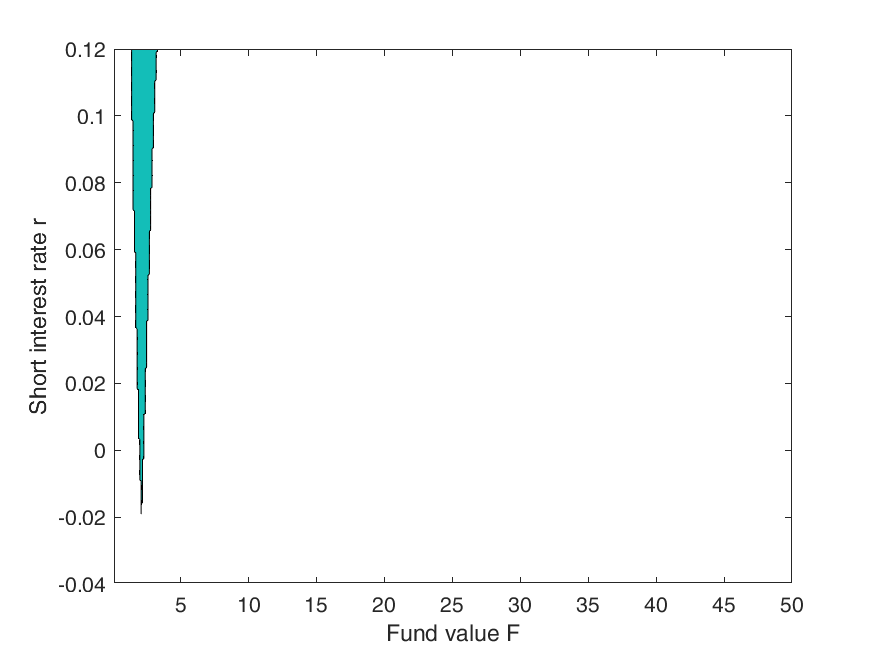}\includegraphics[width=0.35\textwidth]{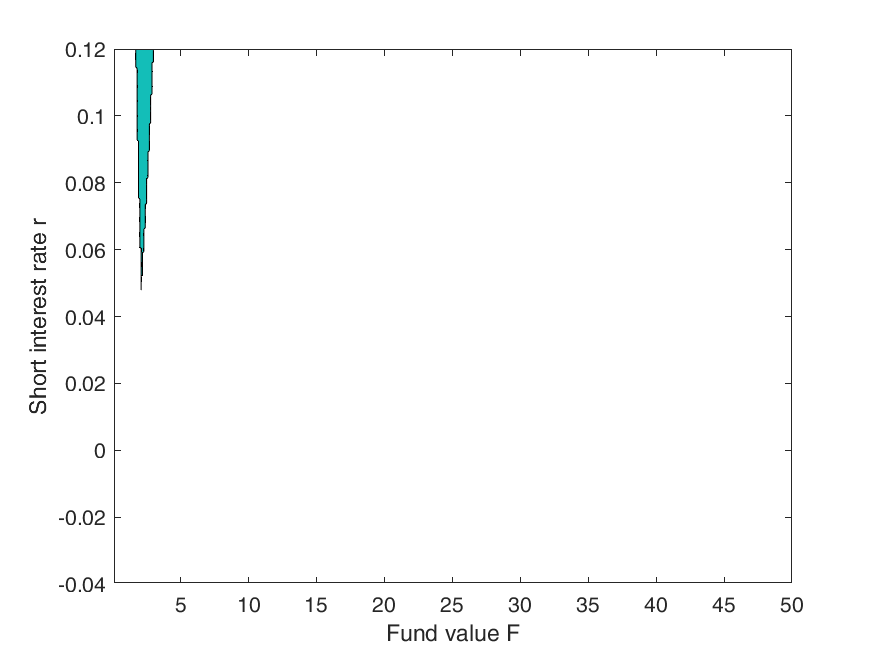}
\par\end{centering}
}
\par\end{centering}
\begin{centering}
\subfloat[$m=10$.]{\begin{centering}
\includegraphics[width=0.35\textwidth]{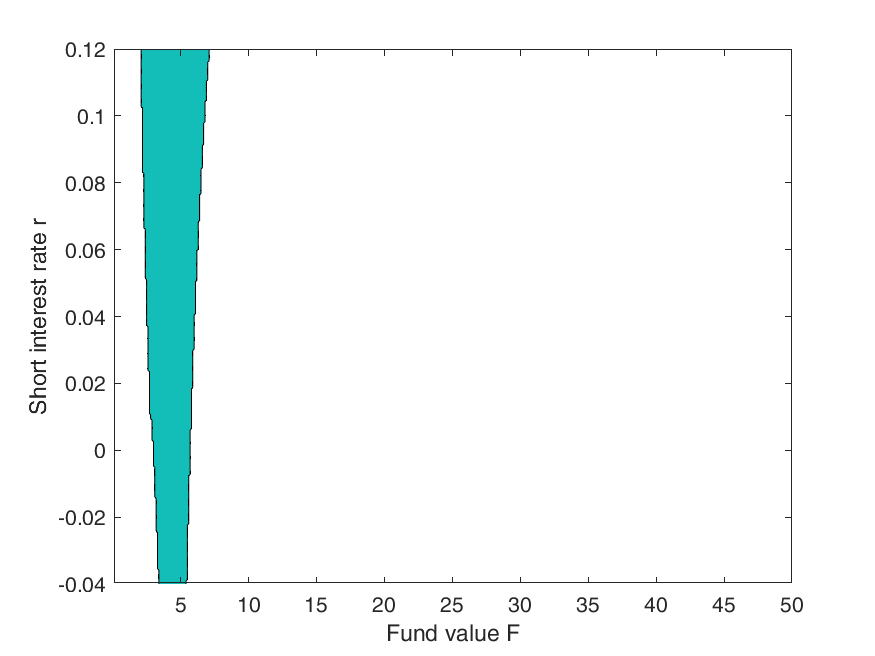}\includegraphics[width=0.35\textwidth]{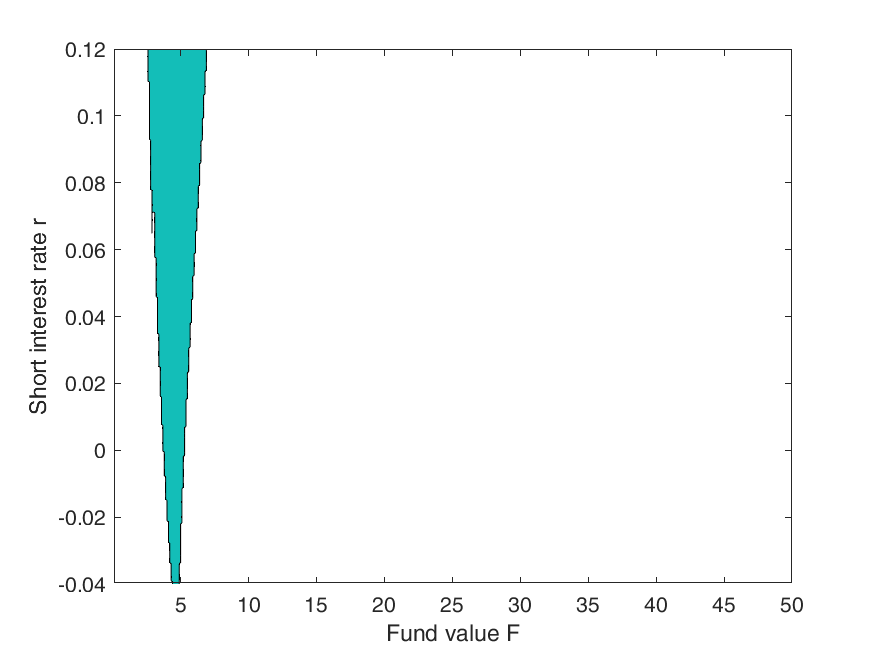}
\par\end{centering}
}
\par\end{centering}
\begin{centering}
\subfloat[$m=15$.]{\begin{centering}
\includegraphics[width=0.35\textwidth]{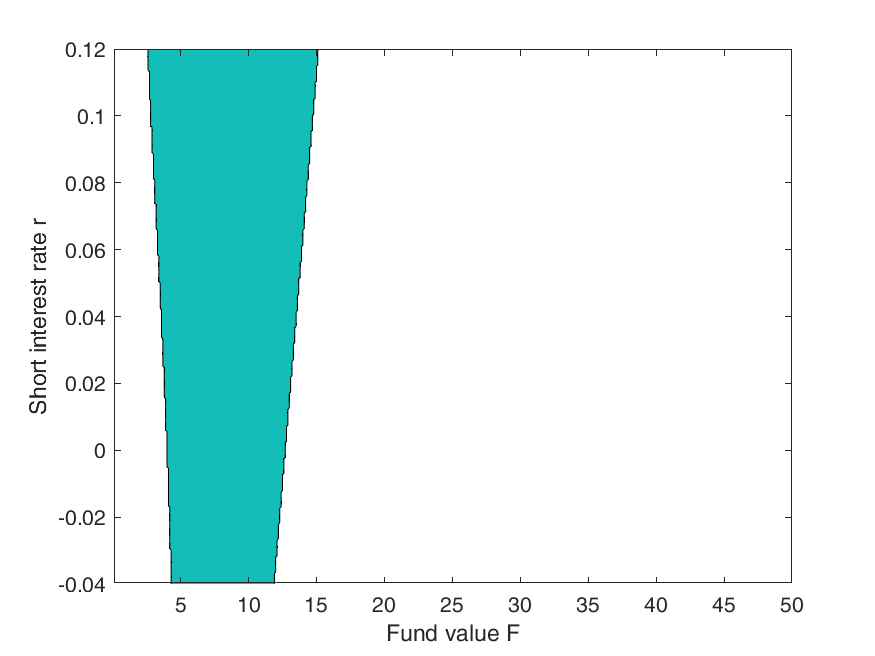}\includegraphics[width=0.35\textwidth]{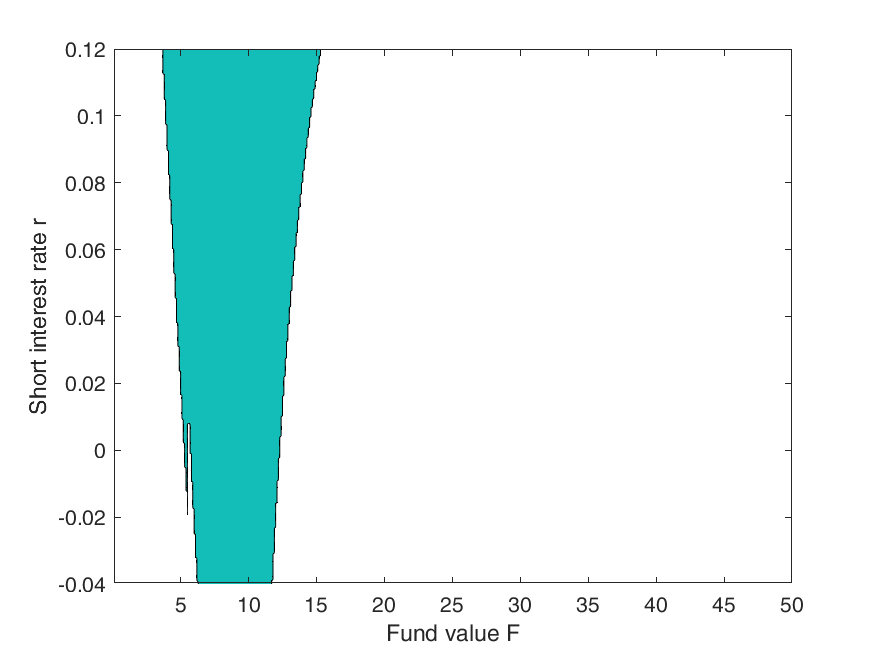}
\par\end{centering}
}
\par\end{centering}
\begin{centering}
\subfloat[$m=20$.]{\begin{centering}
\includegraphics[width=0.35\textwidth]{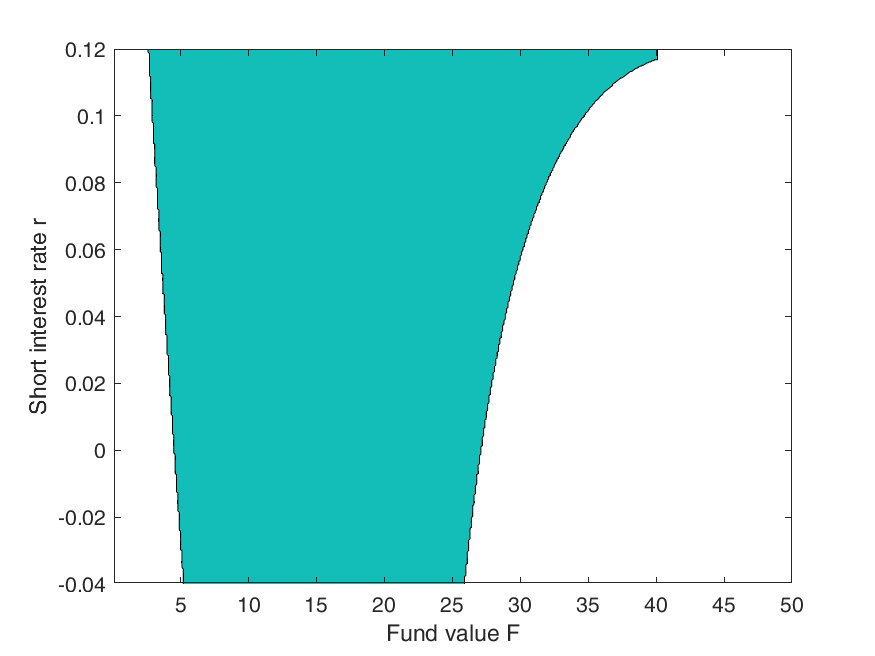}\includegraphics[width=0.35\textwidth]{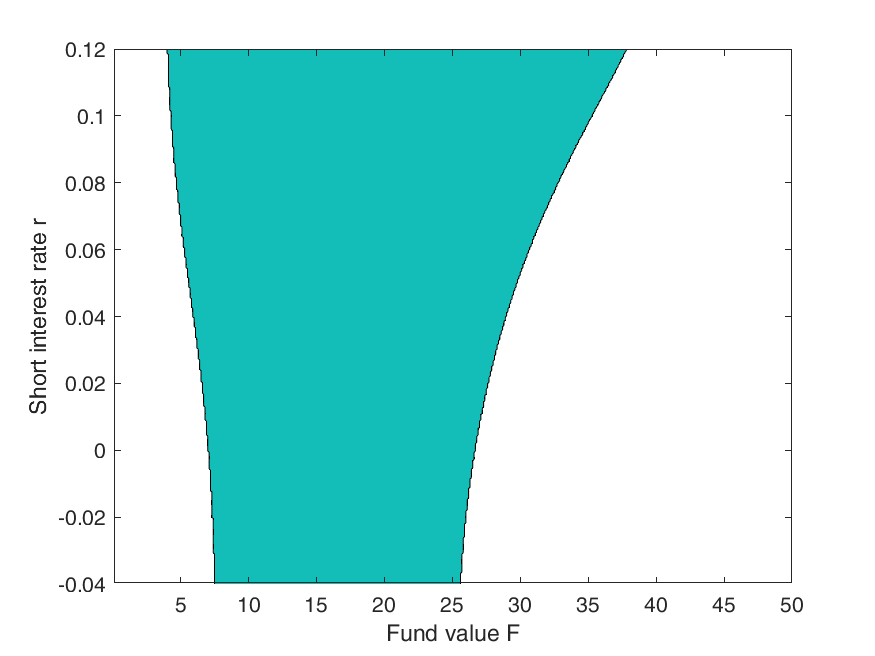}
\par\end{centering}
}
\par\end{centering}
\caption{\label{fig_11} optimal surrender strategy at different anniversaries
for $c=15\%$ and for $g=1\%$ (first column) and $g=3\%$ (second
column). The green area (darker) indicates where it is optimal to
surrender.}
\end{figure}

Finally, in Figure \ref{fig_11}, we plot the optimal early exercise
region, i.e. the values of $F$ and $r$ for which it is convenient
to surrender. We plot these regions for different value of anniversaries
dates, namely $m=5,10,15$ and $20$ for both $g=1\%$ and $g=3\%$.
From the graphs we can observe that in all cases early surrender is
convenient when $F$ takes intermediate values (in-the-money contract),
while it is not convenient when it takes extreme values (out-of-money
contract). Furthermore, the optimal strategy is influenced by the
value of the interest rate $r$: in general, high values of $r$ incentivise
surrender. Consequently, a rise in rates encourages lapse, creating
potential risks for insurers. We can also observe that surrender is
less advantageous in the first few years, and more so as time goes
by: policyholders are less likely to surrender at the beginning, to
benefit from the protection of the contract for a long time, and are
more likely to surrender later to stop paying fees. Finally, it is
worth noting that the strategy that is most convenient for the policyholder
is also the most expensive for the insurer. Therefore, the latter
might find this chart useful in determining when an early termination
of the policy might be more costly for them, based on current assessments
of the fund value and the short-term interest rate. 

\section{Conclusions }

In this paper, we have presented a comprehensive framework for the
valuation of variable annuities within Lévy models, considering stochastic
interest rates. Our approach integrates the robustness of the Hull-White
model to account for the dynamical financial environment and allows
for the incorporation of realistic market conditions into the valuation
process. Through the implementation of advanced numerical methods,
namely the Hybrid method and the Longstaff-Schwartz Monte Carlo method,
we have demonstrated the impact of stochastic interest rates on the
strategic decision-making process related to the optimal surrender
of financial instruments.

Our findings highlight the necessity of considering stochastic interest
rates for long-term financial products and contribute to the deeper
understanding of the risk and return profile of variable annuities.
Moreover, we have shown that our enhanced valuation model can significantly
influence the structuring of contracts, particularly in ways that
disincentivize premature surrender while accommodating realistic market
fluctuations.

The numerical results obtained through our experiments confirm the
validity of the proposed models and open new avenues for future research.
As part of our ongoing work, we aim to explore the broader implications
of stochastic rate models in other areas of financial planning and
risk assessment.

In summary, the insights gained from our study serve as a critical
step towards more sophisticated pricing and risk management strategies
for variable annuities, which are essential to ensuring financial
security and stability for policyholders in an unpredictable economic
landscape. 

\newpage \clearpage

\bibliographystyle{abbrv}
\bibliography{bibliography}

\appendix

\section{Characteristic exponent of Lévy processes\label{LP}}

In this Appendix, we fix the notation for the characteristic exponents
of some of the principal Lévy processes used throughout this document.
The characteristic exponent of a Lévy process, denoted by $\psi(\xi)$,
plays a crucial role in the analysis of these processes.

\subsection{Normal Inverse Gaussian (NIG)}

The Normal Inverse Gaussian process is defined by its characteristic
exponent: 
\[
\psi_{\text{NIG}}(\xi)=\delta_{\text{NIG}}\left(\sqrt{\alpha_{\text{NIG}}^{2}-(\beta_{\text{NIG}}+i\xi)^{2}}-\sqrt{\alpha_{\text{NIG}}^{2}-\beta_{\text{NIG}}^{2}}\right),
\]
 where $\alpha_{\text{NIG}}>0$, $|\beta_{\text{NIG}}|<\alpha_{\text{NIG}}$,
and $\delta_{\text{NIG}}>0$ are parameters of the process.

\subsection{Variance Gamma (VG)}

The characteristic exponent of the Variance Gamma process is given
by: 
\[
\psi_{\text{VG}}(\xi)=-\frac{1}{\kappa_{\text{VG}}}\log\left(1-i\xi\theta_{\text{VG}}\kappa_{\text{VG}}+\frac{1}{2}\kappa_{\text{VG}}\sigma_{\text{VG}}^{2}\xi^{2}\right),
\]
 where $\kappa_{\text{VG}},\sigma_{\text{VG}}>0$ and $\theta_{\text{VG}}\in\mathbb{R}$
are the process parameters.

\subsection{CGMY}

The CGMY process has the characteristic exponent:
\[
\psi_{\text{CGMY}}(\xi)=C_{\text{CGMY}}\cdot\Gamma(-Y)\left[G_{\text{CGMY}}^{Y_{\text{CGMY}}}-(G_{\text{CGMY}}+i\xi)^{Y_{\text{CGMY}}}+M_{\text{CGMY}}^{Y_{\text{CGMY}}}-(M_{\text{CGMY}}-i\xi)^{Y_{\text{CGMY}}}\right],
\]
 with $C_{\text{CGMY}}>0$, $G_{\text{CGMY}}>0$, $M_{\text{CGMY}}>0$,
and $Y_{\text{CGMY}}<2$.

\subsection{Merton Jump Diffusion (MJD)}

The characteristic exponent for the Merton Jump Diffusion process
is: 
\[
\psi_{\text{MJD}}(\xi)=\frac{1}{2}\left(\sigma_{\text{MJD}}\xi\right)^{2}-\lambda_{\text{MJD}}\left(\exp\left(i\mu_{\text{MJD}}^{J}\xi-\frac{1}{2}\left(\sigma_{\text{MJD}}^{J}\xi\right)^{2}\right)-1\right),
\]
 where $\sigma_{\text{MJD}}$, $\lambda_{\text{MJD}}$, $\mu_{MJD}^{J}$,
and $\sigma_{MJD}^{J}$ are the parameters of the MJD process. 
\end{document}